\documentclass[12pt]{article}
\usepackage{amsmath,amsthm,amssymb}
\usepackage[margin=1in]{geometry}
\usepackage[numbers,square,sort]{natbib}
\usepackage{mathrsfs}
\usepackage{booktabs}
\usepackage{nicefrac}
\usepackage{array}
\usepackage{varwidth}
\usepackage{thm-restate}
\usepackage{bbm}
\usepackage{graphicx}
\usepackage{subcaption}
\usepackage{caption}
\usepackage{enumitem}
\usepackage{wrapfig}
\usepackage{cancel}

\usepackage{tcolorbox}
\definecolor{systemcolor}{HTML}{A9A9A9}
\definecolor{usercolor}{HTML}{2980B9}
\definecolor{assistantcolor}{HTML}{27AE60}
\newtcolorbox{systemmessagebox}{
  colback=systemcolor!10, colframe=systemcolor,
  sharp corners, boxrule=1pt, left=5pt, right=5pt, top=5pt, bottom=5pt,
  fonttitle=\bfseries, title=System
}

\newtcolorbox{usermessagebox}{
  colback=usercolor!10, colframe=usercolor,
  sharp corners, boxrule=1pt, left=5pt, right=5pt, top=5pt, bottom=5pt,
  fonttitle=\bfseries, title=User
}

\newtcolorbox{assistantmessagebox}{
  colback=assistantcolor!10, colframe=assistantcolor,
  sharp corners, boxrule=1pt, left=5pt, right=5pt, top=5pt, bottom=5pt,
  fonttitle=\bfseries, title=Assistant
}

\usepackage{tikz}
\usepackage{amsmath}

\usepackage{hyperref}
\usepackage{cleveref}
\usepackage{xcolor}

\newtheorem{theorem}{Theorem}[section]

\theoremstyle{definition}
\newtheorem{definition}[theorem]{Definition}

\Crefname{assumption}{Assumption}{Assumptions}

\usepackage{newunicodechar}
\DeclareUnicodeCharacter{202F}{\,}
\title{Anatomy of a Machine Learning Ecosystem:\\2 Million Models on Hugging Face}
\author{Benjamin Laufer\thanks{Cornell Tech, \texttt{bdl56@cornell.edu}.}\textsuperscript{\ ,\ddag},\  \ Hamidah Oderinwale\thanks{McGill University,  \texttt{hamidah.oderinwale@mail.mcgill.ca}.
}\textsuperscript{\ ,}\thanks{Equal contribution.}, \  \ Jon Kleinberg\thanks{Cornell University.
}}
\date{}
\newenvironment{acks}
{
  \section*{Acknowledgements}
}
{
The authors thank the AI, Policy and Practice (AIPP) group at Cornell University and the Digital Life Initiative (DLI) at Cornell Tech. We thank Samuel J Klein, Daniel van Strien (in-house librarian at Hugging Face), Aleksandra Korolova, Helen Nissenbaum, Tori Qiu, Solon Barocas, Madiha Zahrah Choksi, Gopal Raman, and Manish Raghavan for illuminating conversations and comments. BL also wishes to thank Professor Simon Levin, whose undergraduate course ``Theoretical Ecology'' was an inspiration.

This work is supported by a grant from the John D. and Catherine T. MacArthur Foundation. Ben Laufer is also supported by a doctoral fellowship from DLI, a LinkedIn-Bowers CIS PhD Fellowship, and a SaTC NSF grant CNA-1704527. Hamidah Oderinwale is also supported by a grant from Emergent Ventures and the Carina Fund. Jon Kleinberg is also supported by a Vannevar Bush Faculty Fellowship, AFOSR aware FA9550-19-1-0183, and a grant from the Simons Foundation. BL and HO are additionally supported by a grant from the Cosmos Institute. Any opinions, findings, conclusions, or recommendations expressed in this material do not reflect the views of NSF or other funding agencies.
}

\begin{document}
\maketitle
\begin{abstract}
Many have observed that the development and deployment of generative machine learning (ML) and artificial intelligence (AI) models follow a distinctive pattern in which pre-trained models are adapted and fine-tuned for specific downstream tasks. This process is responsible for much of the powerful functionality we see in current AI systems, but there is limited empirical work that examines the actual structure of these interactions across the AI development community broadly construed. Here we analyze 1.86 million models on Hugging Face, a leading peer production platform for model development. Our study of model family trees---networks that connect fine-tuned models to their base or parent---reveals sprawling fine-tuning lineages that vary widely in size and structure. Using an evolutionary biology lens to study ML models, we use model metadata and model cards to measure the \textit{genetic similarity} and \textit{mutation of traits} over model families. We find that models tend to exhibit a family resemblance, meaning their genetic markers and traits exhibit more overlap when they belong to the same model family. However, these similarities depart in certain ways from standard models of asexual reproduction, because mutations are fast and directed, such that two `sibling' models tend to exhibit more similarity than parent/child pairs. Further analysis of the directional drifts of these mutations reveals qualitative insights about the open machine learning ecosystem: Licenses counter-intuitively drift from restrictive, commercial licenses towards permissive or copyleft licenses; models evolve from multi-lingual compatibility towards English-only compatibility; and model cards reduce in length and standardize by turning, more often, to templates and automatically generated text. Overall, this work takes a step toward an empirically grounded understanding of model fine-tuning and suggests that ecological models and methods can yield novel scientific insights about the development of cutting-edge AI models.
\end{abstract}

\section{Introduction}

Generative artificial intelligence (AI) and machine learning (ML) models are being adopted across a variety of domains. As these technologies develop, there is notable diversity in their levels of availability and paths of diffusion. For example, fully closed-source models may be available through chatbots and API calls, but their weights, source code, training data, and other artifacts remain hidden from view. In contrast, open models make some or all of these materials publicly available for developers and downstream users. 

\begin{figure*}
  \centering
    \hfill
  \begin{subfigure}[t]{0.44\linewidth}
    \includegraphics[width=\linewidth]{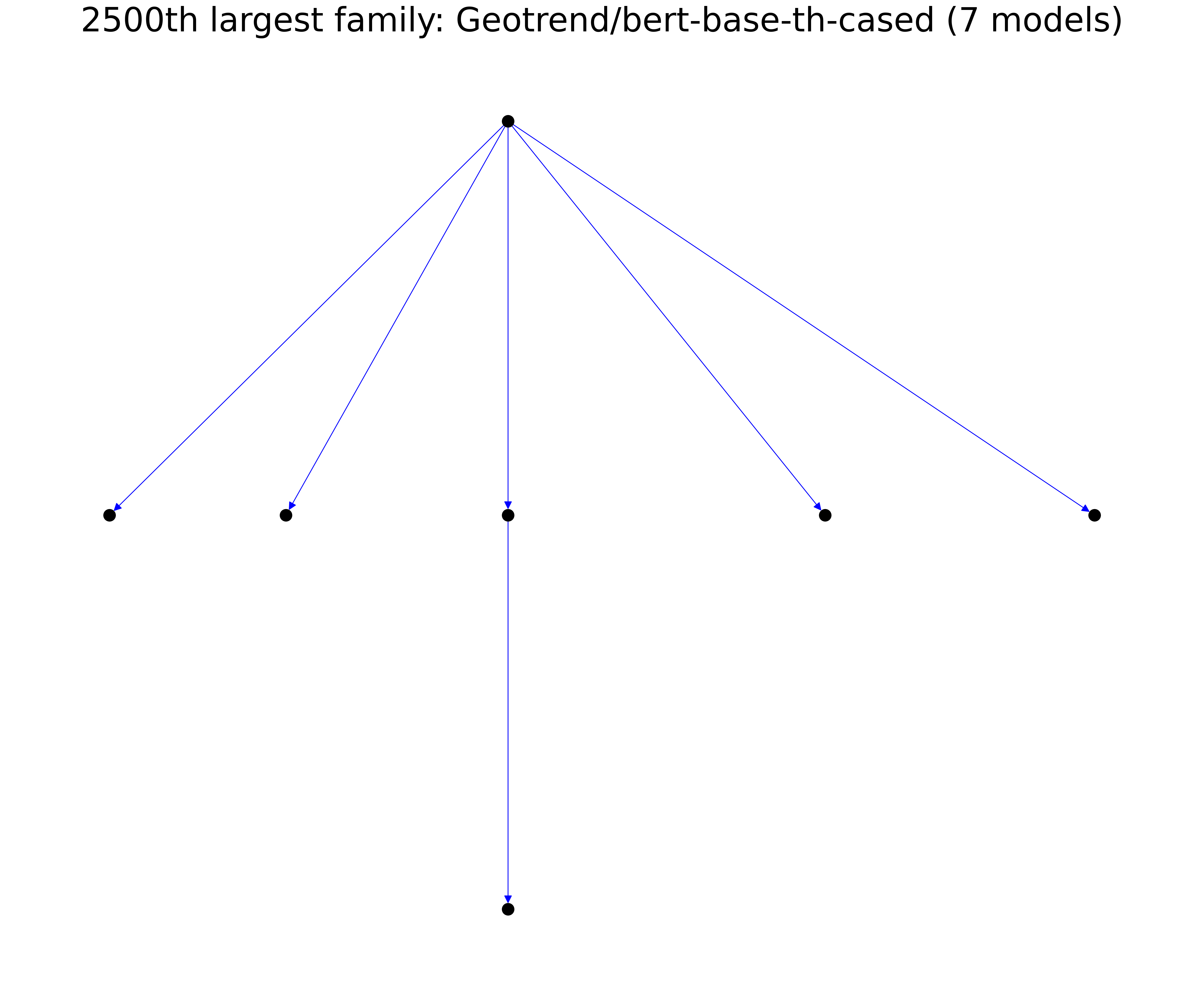}
  \end{subfigure}\hfill
  \begin{subfigure}[t]{0.44\linewidth}
    \includegraphics[width=\linewidth]{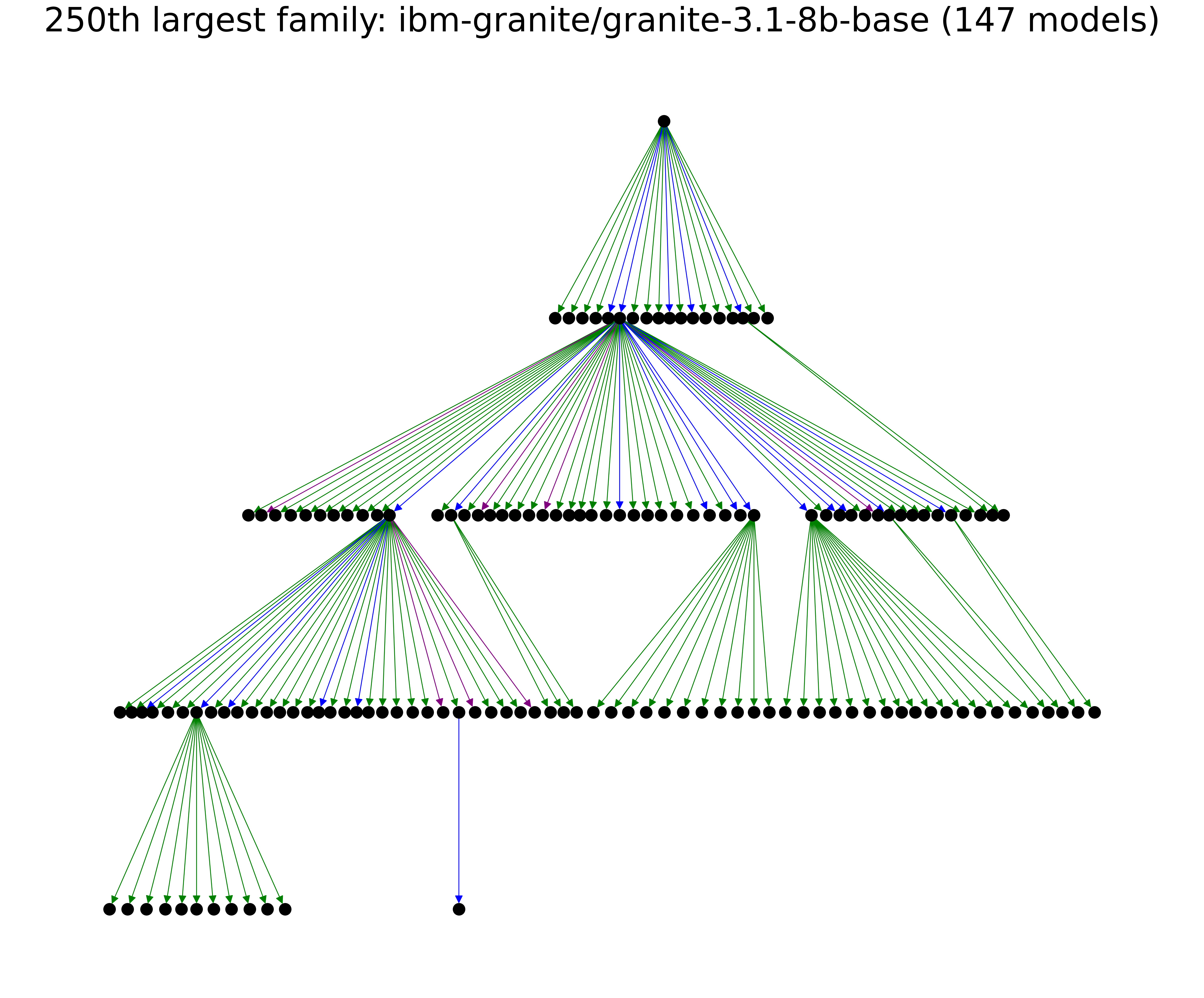}
  \end{subfigure}
\hfill
  \par\smallskip
\hfill
  \begin{subfigure}[t]{0.45\linewidth}
    \includegraphics[width=\linewidth]{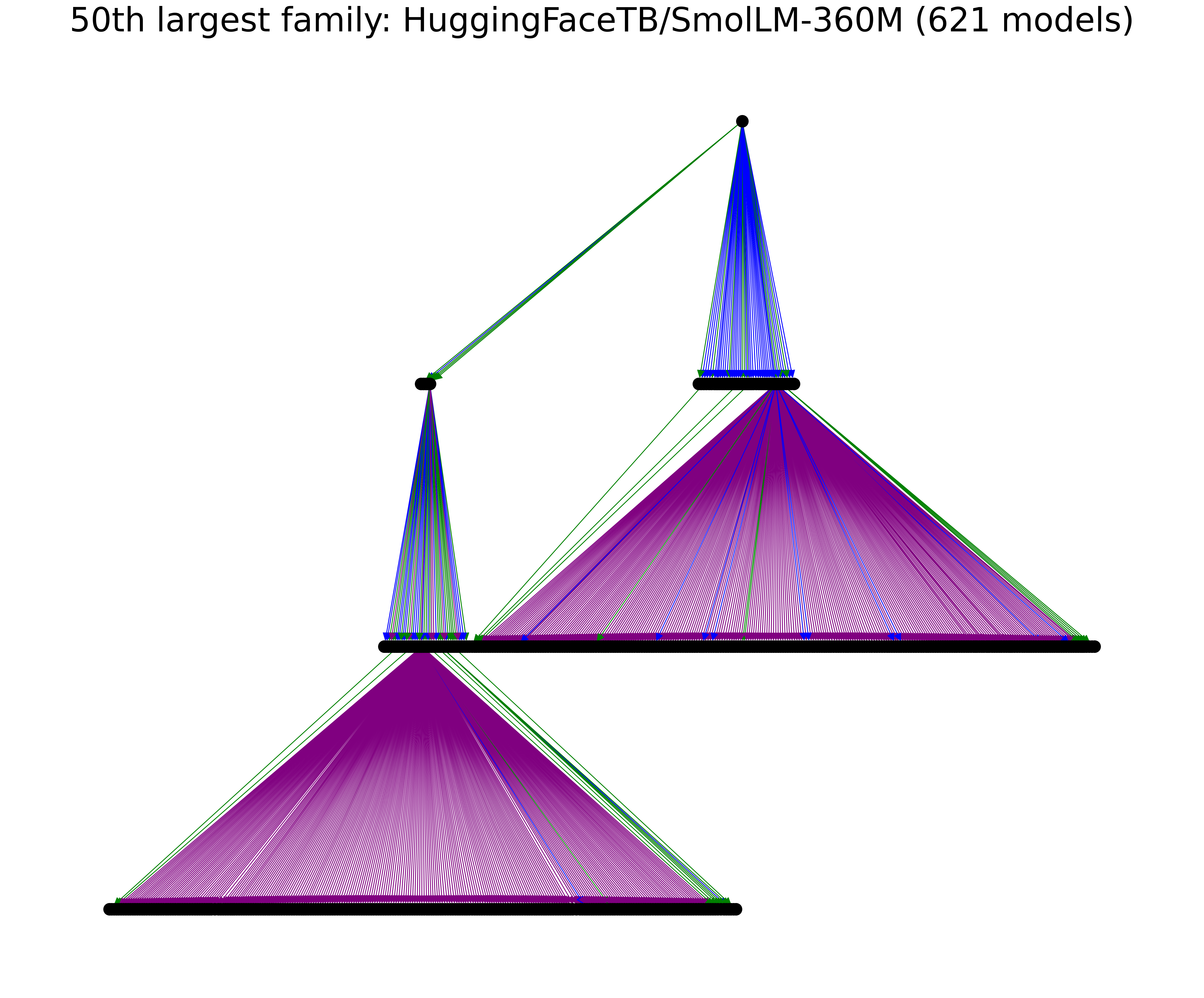}
  \end{subfigure}\hfill
  \begin{subfigure}[t]{0.45\linewidth}
    \includegraphics[width=\linewidth]{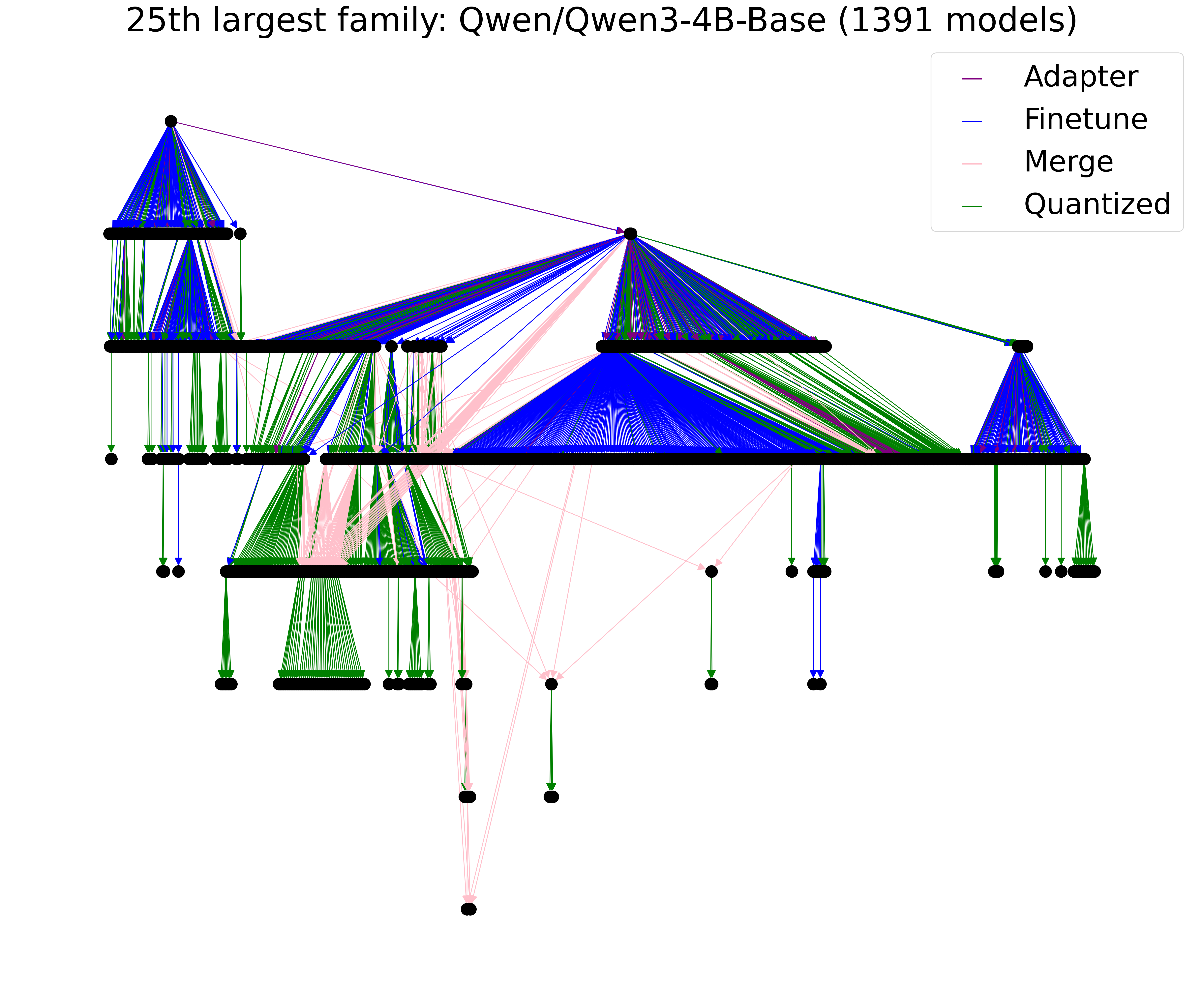}
  \end{subfigure}
\hfill
  \caption{Family trees from the ecosystem dataset. Edges represent different forms of derivative models that are documented as having finetuned, quantized, adapter or merged existing models. Diffusion patterns reveal large broadcasts and numerous generations of derivatives. Graphs without merges are trees, meaning no model has more than one parent (upper left, upper right, and lower left). All graphs are directed and acyclic.}
  \label{fig:splash}
\end{figure*}

The leading provider of open source models is Hugging Face, a platform that hosts models, datasets, libraries, and other materials so that communities of developers can easily use AI models, and perhaps even create derivative products that are useful to others. By making AI models readily available, Hugging Face has cultivated an emerging ecosystem of interacting developers of open source models.

The present work offers an empirical analysis of the open source ML/AI ecosystem on Hugging Face. We analyze information from the 1.86 million models indexed on Hugging Face, finding rich structural diversity in the diffusion of innovation of open source ML. In particular, accessing information about model ``family trees,''\footnote{They are described as model families colloquially and on Hugging Face's webpage corresponding to any available model.} we trace the inheritance and mutation of traits between relatives to understand how models develop and evolve.

Consider, as one example, the model \texttt{ibm-granite/granite-3.1}, a language model developed by the company IBM \cite{granite2024granite} and available on Hugging Face. That model's family tree is depicted in the upper right panel of Figure \ref{fig:splash}. The tree network depicts the connections among 147 models---the base model contributed by IBM and 146 derivative models, many contributed by open source developers. Every black dot represents a distinct, publicly available model, and every directed edge represents a path from a parent model to a derivative model---that is, a finetune, quantization, or adapter. The base model produced by IBM spawned eighteen children, of which five are finetunes and the rest are quantizations. Only two of the children---both finetunes---
spawned further generations of children. Moving down another layer, we can see that one of these children produced many more offspring than the other, and from this ancestor there are further branching generations. As of the time that our snapshot data was ingested, this tree spans five generations, but nothing is stopping developers from continuing to tweak, finetune, and produce new members of this family. The two other trees depicted in Figure \ref{fig:splash} are families with considerably higher numbers of members. They demonstrate some of the diverse structures and rich topologies that arise when we map these model families.

This paper makes the following contributions:
\begin{enumerate}
    \item We propose studying the evolutionary biology of machine learning models to 1) understand the complexities of the development process, 2) analyze model properties and traits and 3) characterize evolutionary trends over time. 
    To that end, we provide the largest dataset to date of the open-model development ecosystem. Using a snapshot of all publicly available models hosted on the platform Hugging Face, our dataset contains a variety of features and, crucially, rich information about the topological relations between models---that is, whether they finetune, merge, or adapt from one or more other models.
    \item We use information from the metadata and model card to track the model's \textit{genetic traits} and measure the \textit{genetic similarity} between models. We find that models of the same family bear a significant resemblance. However, this resemblance departs from typical biological populations because mutations occur at a high rate. For instance, we find that pairs of finetuned siblings share more traits than parent/child pairs, on average, suggesting that mutations occur at a high rate, and are not random but have strong directional trends. 
    \item To understand these directional trends, we conduct a network analysis 
 of the diffusion of model traits between models. We find that mutations of traits including licenses, languages, and tasks are overwhelmingly acyclic; and we solve for optimal orderings over these properties. These orderings allow us to verify, for instance, that \texttt{translation} models are genetically upstream from \texttt{text-generation} models, on whole, and that models with \texttt{llama3} licenses are genetically upstream from those with \texttt{apache-2.0} licenses. The directed and acyclic property of model traits suggests that models are evolving in response to environmental pressures, and we produce a number of hypotheses from the data about the nature of these pressures, which may come from a surrounding market or from behavioral tendencies of developers involved in open source communities. For example, the data suggest that models evolve:
    \begin{enumerate}
        \item From restrictive or commercial licenses toward permissive or copyleft licenses, at times representing a departure from the terms of the upstream model. 
        \item From general multilingual support toward language specialization, with an overwhelming trend toward English-language support.
        \item From expansive documentation practices toward lean or minimal documentation practices.
    \end{enumerate}
\end{enumerate}
These trends yield new hypotheses about the environmental pressures on AI development. For instance, the observation that licenses trend toward permissiveness and copyleft varieties suggests that preferences for open source outweigh existing regulatory pressures to comply with licenses. The drift toward English-language models suggests a formidable market for english-language products. These and other hypotheses are discussed along with future directions for inquiry. 
    
\subsection{Related work}

This paper aims to measure and analyze the structure of AI \textit{fine-tuning} and related adaptation and transfer learning procedures. These relationships connect finetuned and remixed AI models to their `parent' model(s) whose weights, structures, and other elements might influence the child's development. The sources of inspiration for this work come from scholarship on \textbf{social networks and the web}, \textbf{multi-agent interactions and modeling AI development}, and finally, \textbf{approaches from theoretical ecology and genetics}. We cover relevant work from each of these categories in turn.

\textbf{Social Networks on the Web}. 
\citet{goel2016structural} differentiate broadcast diffusion trees from viral trees using a metric they term structural virality, which we use to measure the connected components in Appendix \ref{app:structural-virality}. Many have considered the dependence of graph features on local network topology, including in the context of attachment \cite{ugander2012structural}, link prediction \cite{liben2003link,leskovec2010predicting}, feature prediction \cite{grover2016node2vec,hamilton2017representation} and community inference \cite{gibson1998inferring}. In contrast, our approach attempts to predict trait similarity and trait transitions over a tree network. Though empirical work on Hugging Face is limited, some strides have been made. \citet{horwitz2025charting} calls for work mapping an `atlas' of models on Hugging Face, demonstrating that directed acyclic graphs representing model relationships can be drawn for certain families and providing a dataset with 1.1 million models. Our work answers this call and offers an expanded dataset. \citet{choksi2025brief} explore chats and conversations among community members and contributors, evidence of vibrancy and richness among contributing developers. \citet{bommasani2023ecosystem} coin \textit{ecosystem graphs} as an abstraction for understanding AI development, and analyze a preliminary set of 128 models that they use to demonstrate the usefulness of ecosystems thinking for reasoning about social implications and regulation of AI. \citet{duanposition} tracks the frequency of copyleft license violations across model derivatives using a dataset of around 15,000 models on Hugging Face. \citet{rahman2025hugginggraph} use the Hugging Face API to create a graph of information about models totaling 402,654 nodes.

\textbf{Multi-agent interactions and modeling}. Scholars have developed theoretical models and theories of the multi-actor system surrounding the development of AI technologies. \citet{laufer2024fine} create a game-theoretic model to understand how `domain specialists' and `generalists' interact to produce the technology. Others have developed depth-one tree structures as a model for understanding AI diffusion \cite{jagadeesan2024safety,qiu2025formal,dean2024accounting,laufer2025backfiring}. \citet{hopkins2025ai} use directed acyclic graphs (DAGs) of arbitrary depth to allow supply chains of interacting actors to understand the dynamics of AI supply chains. There is budding work on decision-making along these networks \cite{widder2023dislocated, taitler2025selective}, though much of it is theoretical. Further, we claim that perspectives on incentives, competition, cooperation have tended to be organized by economic---rather than ecological---metaphors. Here, we wish to go deeper with the ecological phenomenology of AI development and diffusion.

\textbf{Theoretical Ecology and Genetics}. 
This paper is inspired by perspectives of systems as \textit{complex adaptive systems}, characterized by emergent properties that arise from small-scale interactions between components \cite{levin1998ecosystems}. \citet{sclocchi2024phasetransitiondiffusionmodels}, taking a machine learning perspective, understand model `phylogeny' as a prediction problem, and show that models with larger normed parameter vectors---weights and biases of greater magnitude---tend to be higher up in the family tree. In a different genealogical approach to machine learning, \citet{kalluri2025computer} draw links between ML papers and downstream produce developments, focusing on surveillance applications.

\section{A dataset of 1.86 million models on Hugging Face}

In this work, we examine the Hugging Face model hub, the largest public repository of machine learning models, containing 1.86 million models at the time of this study. 
We approach Hugging Face as a platform for peer production, building on prior research into collaborative systems and the structure of the broader web \cite{benkler2006commons, 10.5555/1784297.1784348, 10.1145/276627.276652}. 
With rich textual information containing information about the relationship between models as well as their traits, we can represent the complex network of models on Hugging Face as a set of phylogenies—branching trees rooted in base models, where nodes correspond to individual models and edges denote parent–child relationships.\footnote{Our dataset is publicly available at the following link: \href{https://huggingface.co/datasets/modelbiome/ai\_ecosystem\_withmodelcards}{Hugging Face dataset}. Our codebase is available at the following link: \href{https://github.com/bendlaufer/ai-ecosystem/tree/main}{GitHub repository}.}

\subsection{Data collection}

We collected the data for our dataset in two stages. In the first stage, we used the Hugging Face `model' API to collect the model features and relationships---that is, all pieces of information in our dataset aside from the model cards. Hugging Face provides API access to individual \textit{lists} of models, but these lists are capped to only list 1000 models. Using pagination, we were able to iterate over all such lists of models to collect the information in our dataset in JSON format. In the second stage, we collected the full text of every model's model card through individual, per-model API calls to the model cards API. These cards were significantly more data-intensive---since model cards can be quite large and many more API calls were required to find all 1.86 million models in the dataset. In total, our full dataset uses memory on the order of 10GB (depending on the file format used), and the dataset without model cards uses significantly lower memory, at around 500MB. All calls to the API were conducted through the authors' registered accounts on Hugging Face, and in consultation with employees at Hugging Face, including Hugging Face's in-house librarian. 

\subsection{Properties and summary statistics}

Our dataset centers around snippets of text for every model known as the model's metadata. Model metadata comes in JSON format, and this JSON is made readily available for any model through Hugging Face's API. These JSONs include the \textbf{\texttt{model\_id}} (a unique identifier for each model containing its author and name), \textbf{\texttt{likes}}, \textbf{\texttt{trendingScore}} (a trait defined by Hugging Face for ranking models on their website), \textbf{\texttt{downloads}}, \textbf{\texttt{pipeline\_tag}} (also known as \textbf{\texttt{task}}---a categorization of models into e.g., \texttt{feature-extraction}, \texttt{text\--generation}, \texttt{image-classification}, and other modalities), \textbf{\texttt{library\_name}} (the Hugging Face library used to support development), \textbf{\texttt{createdAt}} (the date and time that the model was created\footnote{Tracking of the \texttt{createdAt} date and time began March 2, 2022. According to the Hugging Face documentation, and corroborated by our findings, all models created before that date are back-filled with that date; the date is accurate for all models uploaded thereafter.}), and \textbf{\texttt{tags}}. Tags contain a structured list of strings, some with organized prefixes. For example, tags beginning with \texttt{base\_model:finetune:} link a finetuned model to its parent's model id, tags beginning with \texttt{license:} contain the model's license, and those beginning with \texttt{arxiv:} contain links to the arXiv identifiers of accompanying papers. Other tags do not have these prefixes, but their meaning can still be inferred. For example, languages are listed using two- or three-letter ISO-639 codes. 

\begin{figure*} 
    \centering
    \includegraphics[width=0.9\linewidth]{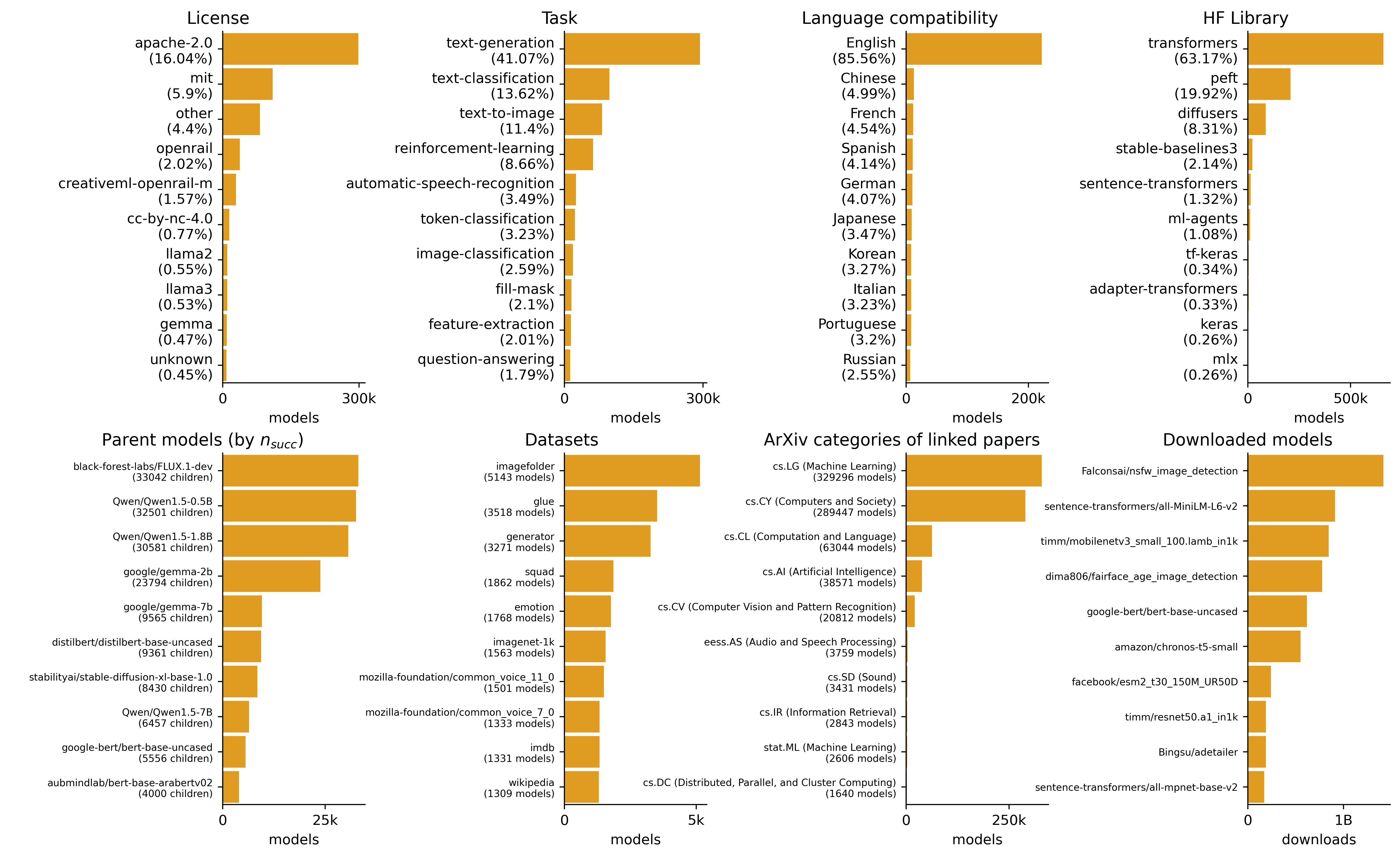}
    \caption{Top ten most frequent licenses, tasks, languages, and libraries (top row). Top ten models ranked by number of children, datasets, arXiv categories of linked papers, and downloaded models (bottom row).
    }
    \label{fig:mega-figure}
\end{figure*}

A summary of the distributions of the various metadata traits is provided in Figure \ref{fig:mega-figure}. These distributions convey the relative frequencies of different traits, as well as the absolute number of papers with these documented traits. Here, we provide some findings these figures convey about the state of the open source ecosystem on Hugging Face, reading from left to right and top to bottom through the figure. 
First, permissive licenses---especially \texttt{apache-2.0} and \texttt{mit} are dominant, constituting over 60\% of all reported licenses. Text-based tasks---and especially \texttt{text-generation}---are most common. English is by far the dominant language compatibility on Hugging Face, with over 75\% of models that document any language compatibility marking english as a supported language. Chinese is the second most-common at 4.4\%. \texttt{transformers} is the most common Hugging Face library. \texttt{black-forest-labs/FLUX.1-dev} is the model that has the most children. \texttt{imagefolder} is the most commonly recorded dataset in metadata.  \texttt{Machine Learning} and \texttt{Computers and Society} codes are the most common among linked arXiv papers. Finally, in the lower right figure, we show the most downloaded models, finding that the model \texttt{Falconsai/nsfw\_image\_detection} is the most downloaded. This model's purpose is to detect and identify explicit imagery and is perhaps used for content moderation and compliance. 

A remarkable amount of information is conveyed in text snippets that Hugging Face stores for every model. Throughout the paper, we treat the snippets of text provided by the metadata JSON as the models DNA, as it contains rich information about traits and allows us to track changes and differences over generations (illustrated in Figure \ref{fig:dna}). Before embarking on this genetic analysis, we discuss one additional source of genetic information: the model cards.

\subsubsection{Model cards} Model cards are documents that carry information about the use, performance, compatibilities, risks, impacts, and many other pieces of information about models \cite{mitchell2019model}. Model cards are the main form of documentation for models on the hub, and they constitute much of the information that populates on any given model's associated webpage. Model cards can be considerably longer than metadata, and much less structured. They can therefore contain more information, however, not all models have corresponding model cards, and they are considerably less standardized and organized. According to our data, 67.04\% of models currently have an associated model card. An analysis of the 1,247,149 cards available reveals an average model card length of 3575.60 characters ($\approx$ 436.06 words), with a median of 2073.0 characters ($\approx$ 238.0 words). This wide range, from a minimum of 11 characters to a maximum of 18,289,454 characters ($\approx$ 2,813,762 words), indicates that a small number of extremely verbose cards significantly influence the average. 

\begin{figure*} 
    \centering
    \begin{minipage}[t]{0.42\linewidth}
        \vspace{0pt}
        \includegraphics[width=\linewidth]{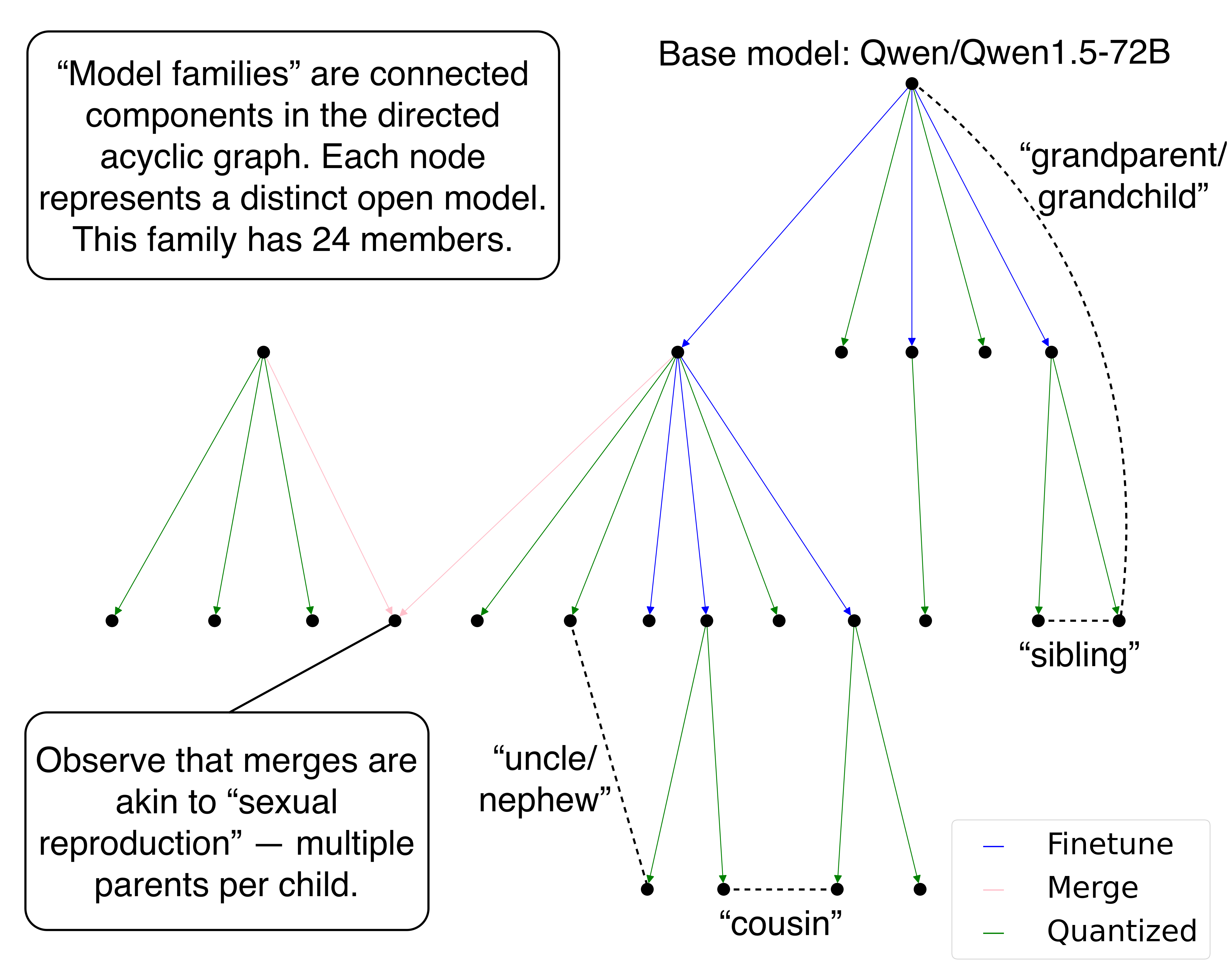}
        \caption*{}
    \end{minipage}
    \hfill
    \begin{minipage}[t]{0.55\linewidth}
        \vspace{0pt}
        \includegraphics[width=\linewidth]{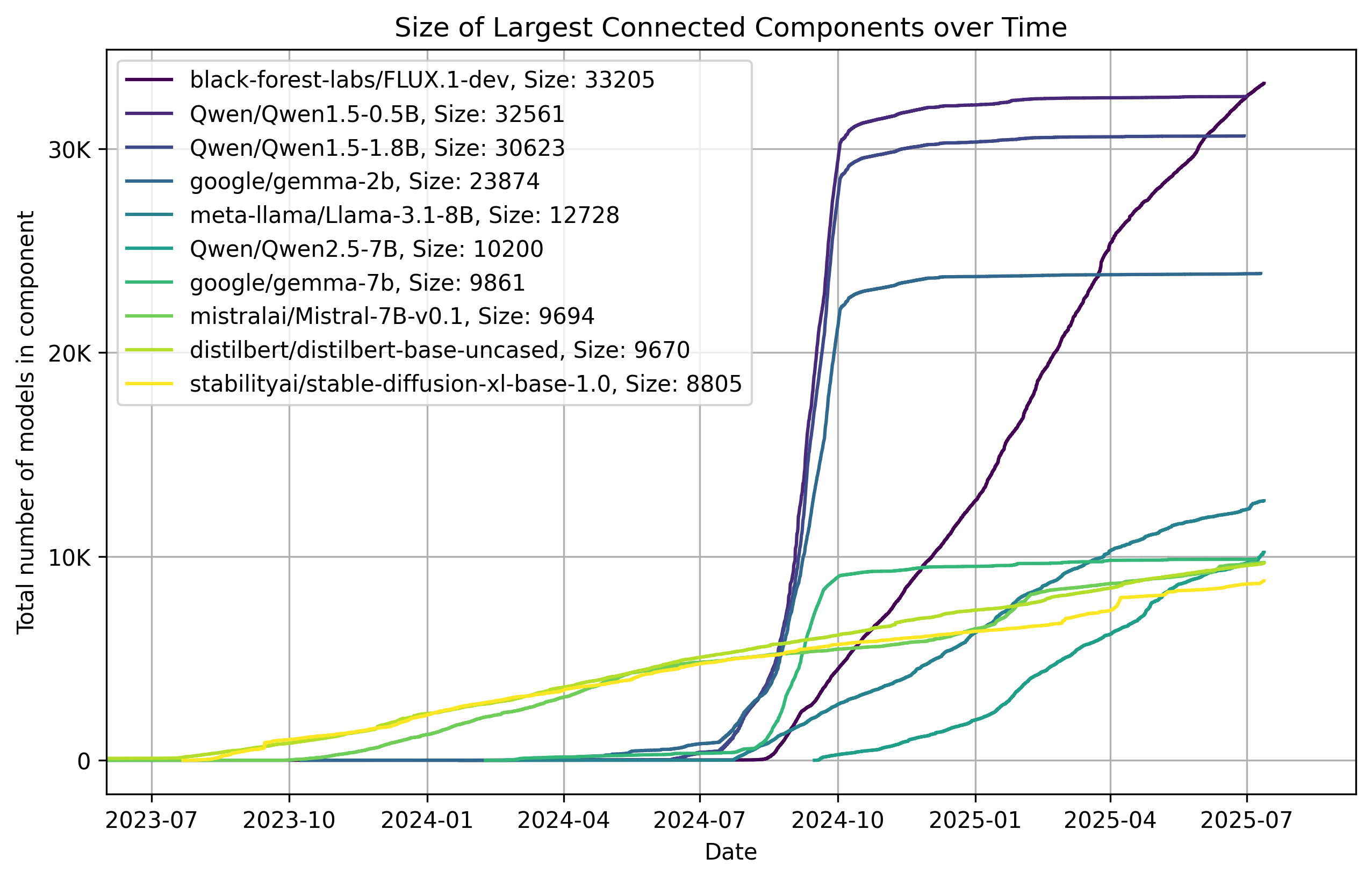}
        \caption*{}
    \end{minipage}

    \caption{Schematic representing different family relationships using an example family tree from our dataset (left). For the ten largest family trees, we depict the growth of the graph over time using the \texttt{CreatedAt} field logged when a new model is created on Hugging Face (right). The growth of the ten largest connected components in our dataset reveals “S-curve” adoption patterns \cite{07951efe-fc63-3730-8cb4-922d670206dc}, analogous to other domains with diffusion over a network.}
    \label{fig:growth-over-time}
\end{figure*}

\begin{figure*}
    \centering
    \includegraphics[width=\linewidth]{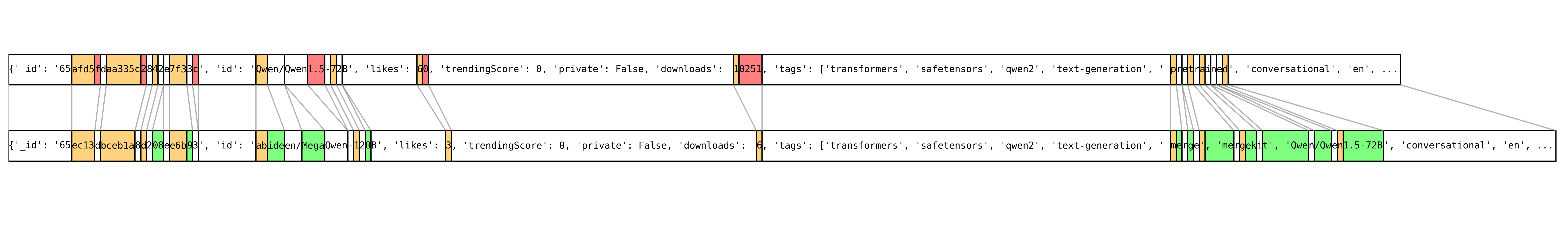}
    \caption{The \texttt{diff} between two sequences of model metadata. We measure the overall mutation rate and genetic similarity by tracking rates of overlap and departure between these sequences. The metadata sequence depicted on top is that of \texttt{Qwen/Qwen1.5-72B}, the base model depicted in Figure \ref{fig:growth-over-time}; the bottom sequence is one of its finetunes. Additions are shown in green, deletions in red, and substitutions in yellow. This figure depicts character-level mutations corresponding most closely to the Levenshtein distance. We additionally measure and report similarity on term-level representations (using bag-of-words and TF-IDF), which we believe better captures categorical shifts in metadata.
    }
    \label{fig:dna}
\end{figure*}

\section{Measuring genetic similarity}

With rich structured data about the relationships between AI models, there are a number of questions we can ask about the diffusion of model attributes. Inspired by ecological and genetic perspectives \cite{hamilton1964genetical, eberhard1975evolution} and existing work on network diffusion \cite{ugander2012structural}, here we explore the relationship between \textit{family structure} and \textit{attribute similarity}. We develop a method for measuring how related two models tend to be given their proximity in our graph. If finetuning family trees are akin to genetic family trees, we might expect two models finetuned from the same parent model (`siblings') to be more similar, on average, than any two models selected at random from our dataset. Taking the metaphor further, if we think of the encodings of model attributes---including licenses, tags, text data, and other metadata information---as akin to DNA in biological species, reproductive models would predict that parent-child pairs tend to be more genetically related than uncle-nephew pairs or grandparent-grandchild pairs.

\textbf{Semantic similarity as genetic similarity.} In living organisms, genes are encoded in a semantic language through sequences of nucleotide bases---or ``building blocks''---in DNA. One way of measuring genetic relation is by measuring the overlap or similarity in DNA sequences. AI models encode their own forms of semantically meaningful instruction sequences through their code bases, model cards, metadata, and model weights. Luckily, open models on Hugging Face make many of these resources publicly available, enabling formal approaches to reasoning about model similarity. Each of these artifacts is different in kind, and of course, none are perfect analogies for DNA. Here, we provide a method for measuring genetic similarity between models, inspired by the genetic metaphor. Our approach measures the semantic distance between the models' tokenized metadata. We propose measuring the frequency of different terms in the model metadata and tracking differences in these relative frequencies.

\begin{figure*} 
    \centering
    \begin{minipage}[c]{\linewidth}
    \centering
    \includegraphics[width=\linewidth]{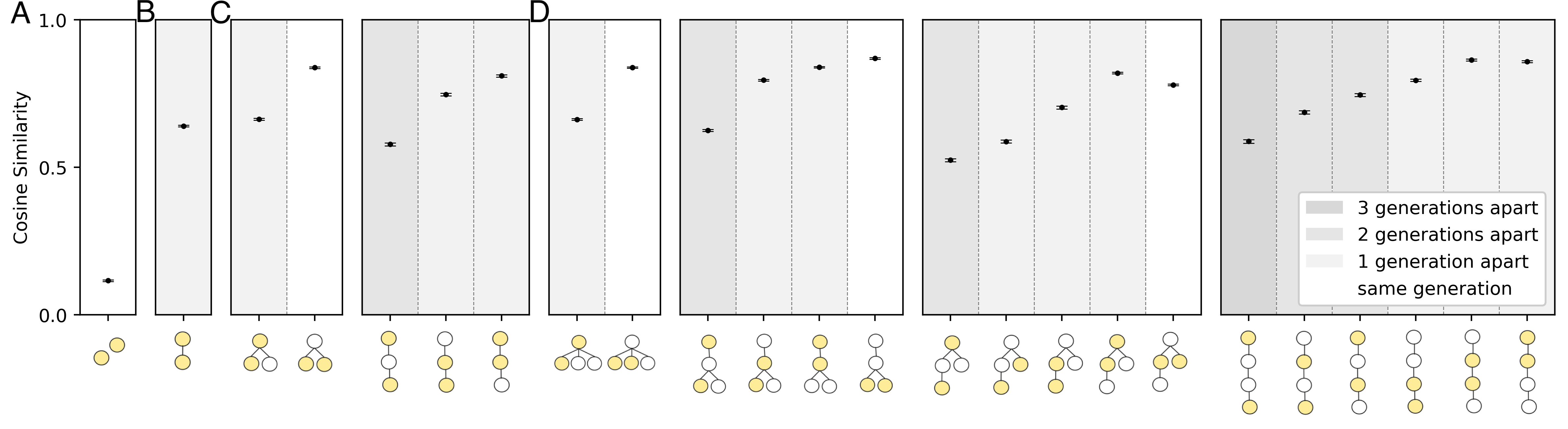}
    \caption{Cosine similarity between TF-IDF embedding vectors, trained on terms appearing in the model metadata for all models in our dataset. Here, we sample finetunes meeting specific family structures. We enumerate all possible sub-trees of size 2 (B), 3 (C), and 4 (D), and enumerate all possible pairs of nodes within these sub-trees. When we compare these genetic similarities to the baseline of the similarity between any two nodes in the graph (A), we find that all observed family ties strongly predict attribute similarity. Similarities between pairs of models suggest that models are more related when they reside at similar depths and when they are topologically close in distance.}
    \label{fig:tfidf-metadata}
    \end{minipage}
    \vspace{1cm}
    \vfill
\centering
\begin{minipage}[c]{0.545\linewidth}
\vfill
\includegraphics[width=\linewidth]{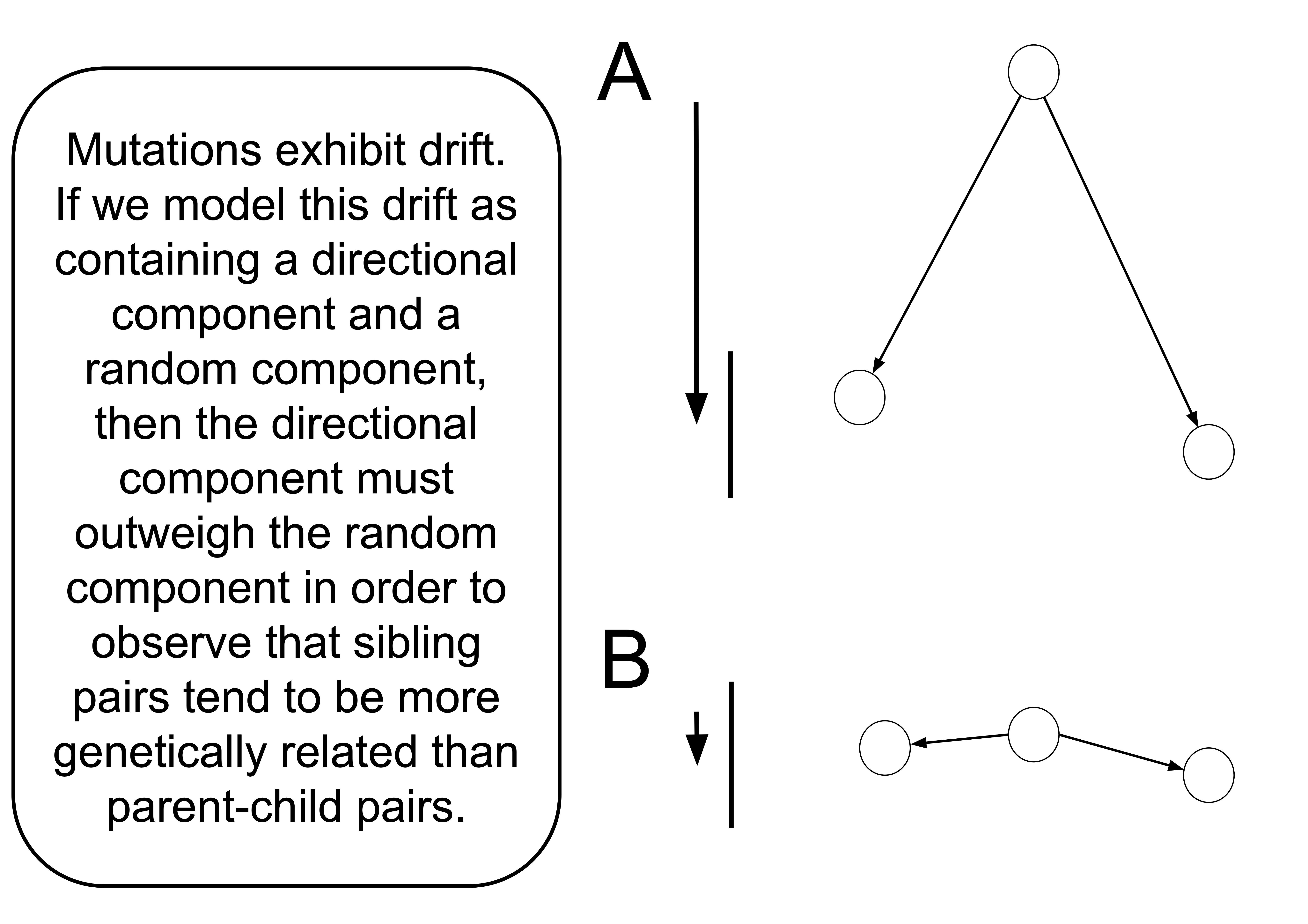}
\vspace{0.66cm}
\vfill
    \caption{We observe that siblings exhibit greater similarity in traits than parent-child pairs. This implies not only that there is a high rate of mutation, but that mutations are sufficiently directed. 
    }
    \label{subfig:mutation-illustration}
\end{minipage}
\hfill
\begin{minipage}[c]{0.425\linewidth}
\centering
\begin{tabular}{cc}
\toprule
{\textbf{Topology}} & {\textbf{Occurrences}} \\
\midrule
\includegraphics[width=0.45cm
]{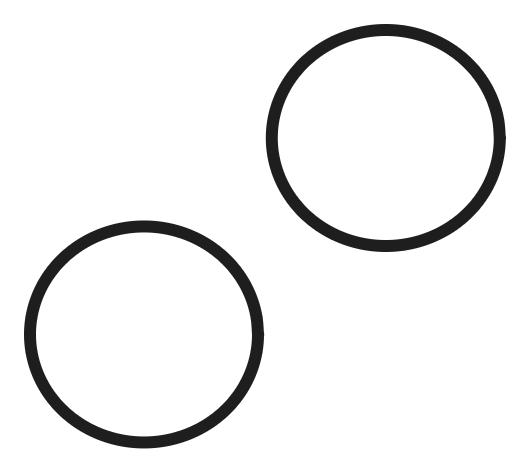} & {3,470,193,356,870} \\
\includegraphics[width=0.25cm
]{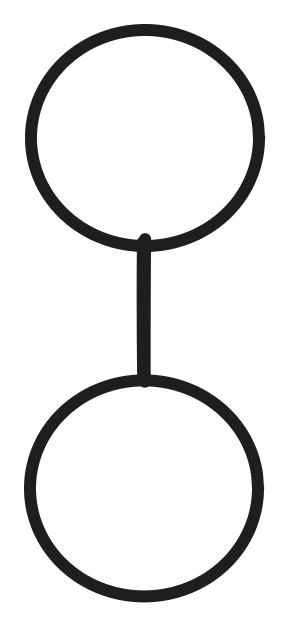} & {191,072} \\
\includegraphics[width=0.45cm
]{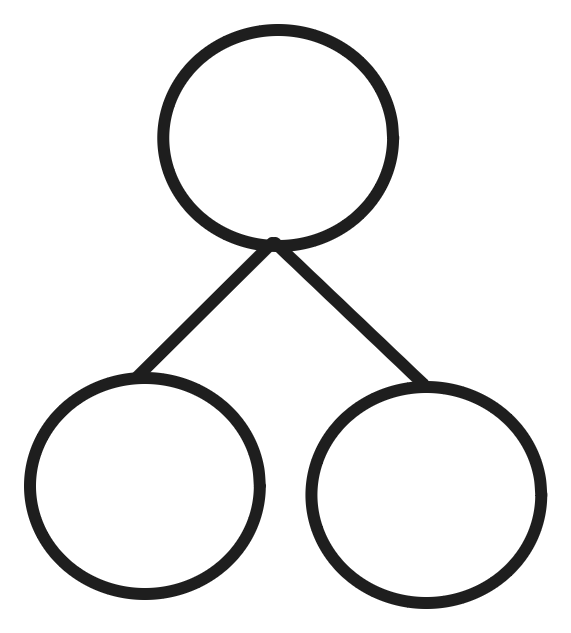} & {119,795,843} \\
\includegraphics[width=0.25cm
]{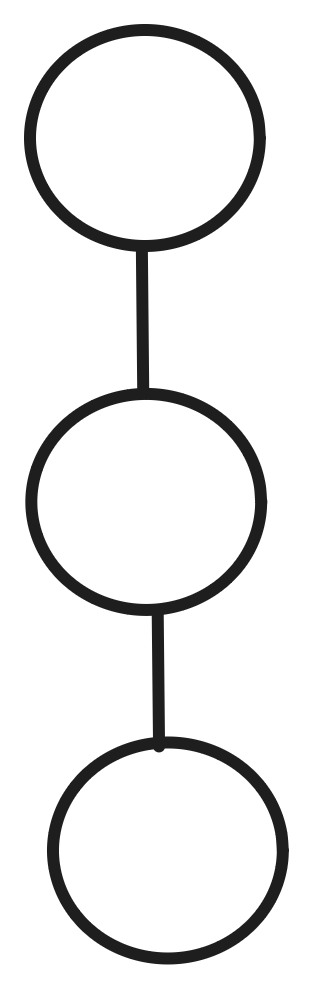} & {40,922} \\
\includegraphics[width=0.6cm
]{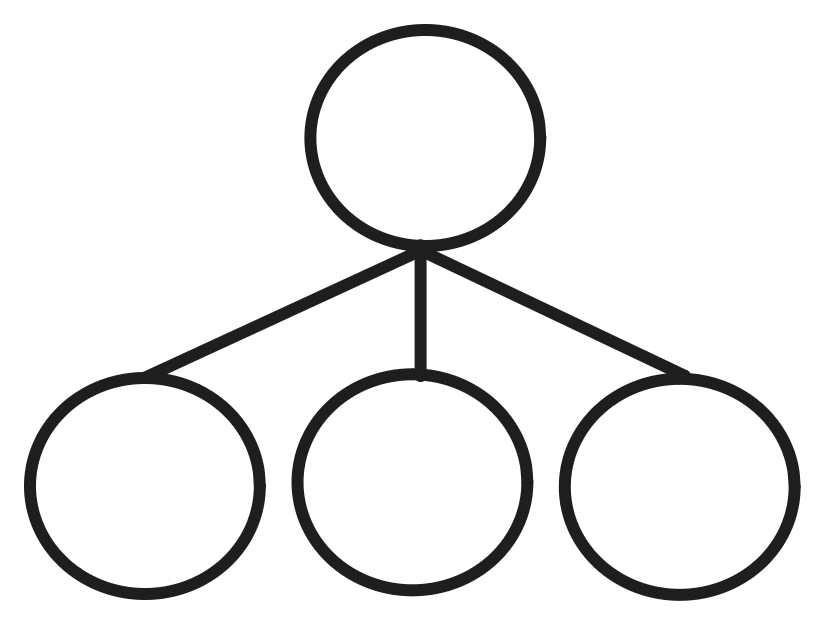} & {193,010,561,824} \\
\includegraphics[width=0.45cm
]{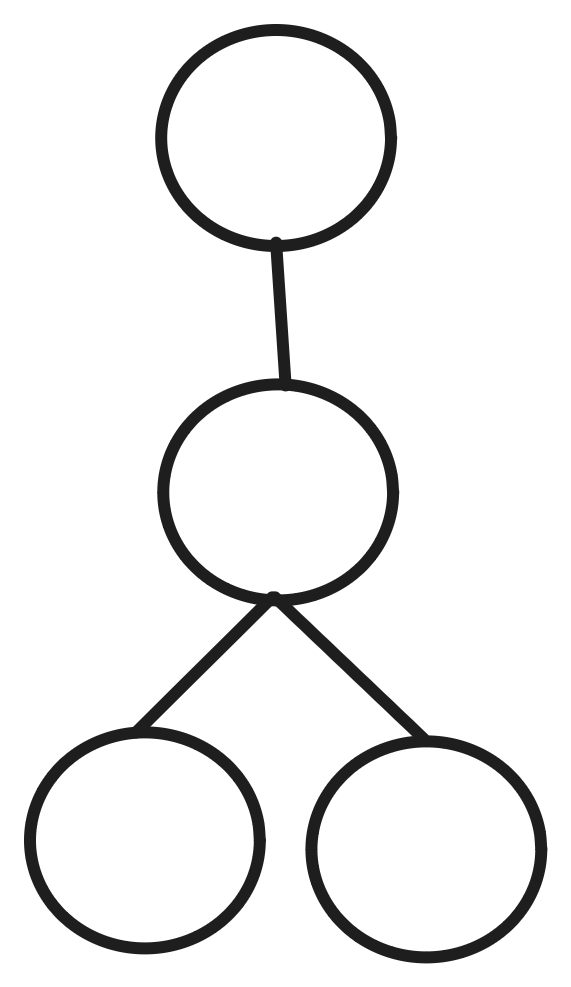} & {11,847,103} \\
\includegraphics[width=0.45cm
]{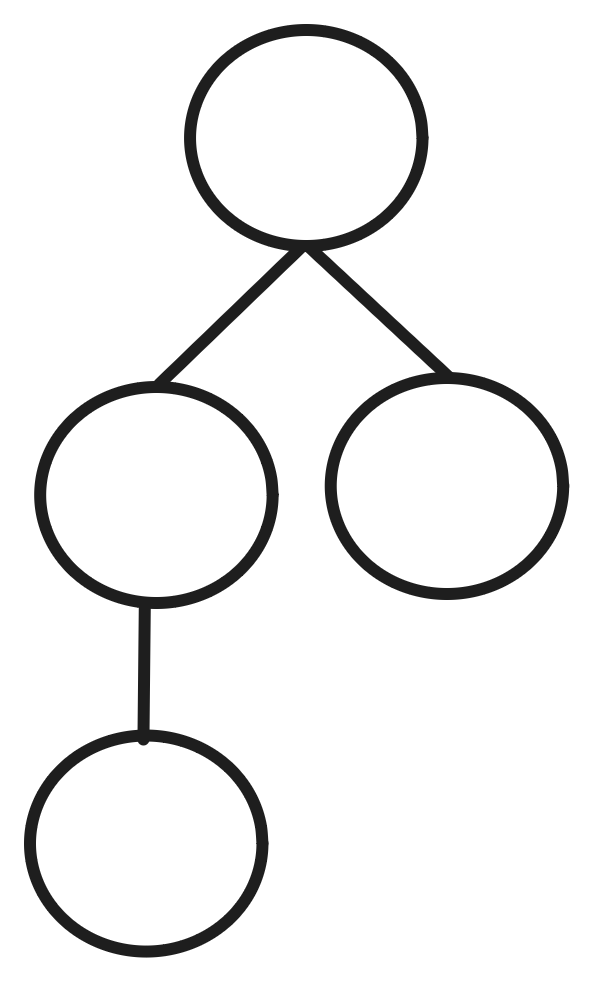} & {19,932,645} \\
\includegraphics[width=0.25cm
]{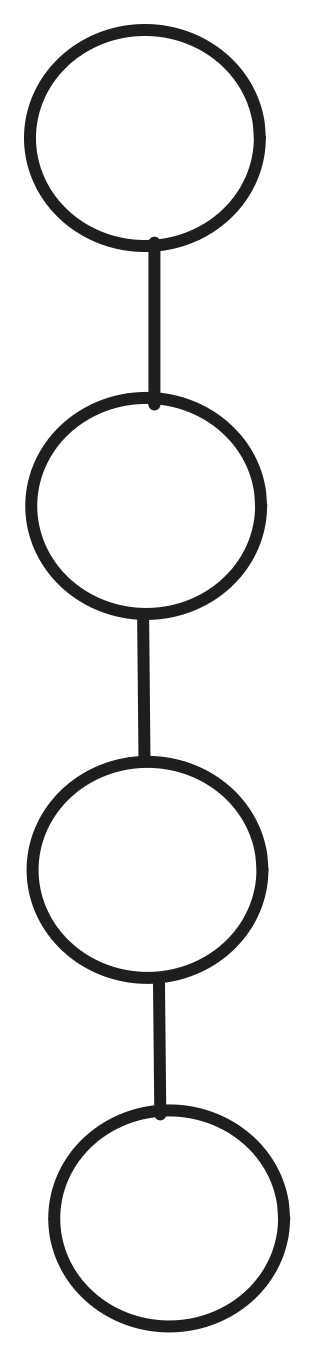} & {10,965} \\
\bottomrule
\end{tabular}
\caption{The graph contains many instances of some family subtrees. Pairwise similarities within subtrees are estimated via sampling.}
\label{tab:occurences-table}
\end{minipage}
\vspace{1ex}
\label{fig:table-figure-combo}
\end{figure*}

Our approach to calculating similarities borrows from classical contributions in natural language processing based on term frequency. We replicate our analysis for three similarity measures -- the normalized Levenshtein Distance \cite{yujian2007normalized}, which directly computes character-level insertions and deletions as depicted in Figure \ref{fig:dna}, the cosine similarity in term frequency (or ``bag-of-words'') embeddings, and the cosine similarity in term frequency-inverse document frequency (``TF-IDF'') embeddings. We measure similarities across two different model artifacts --- metadata JSONs and the text of model cards. Our results across these different metrics reveal the same insights: that models of the same model family are more genetically similar than randomly paired models, and that genetic similarity is negatively related to the generational divide and topological distance. Further information on these various metrics and approaches are provided in Appendix \ref{app:measures}. In the body of the text, we report the cosine similarity on TF-IDF embeddings derived from the metadata strings (Figure \ref{fig:tfidf-metadata}). 

\textbf{Analyzing fine-tuning trees.} 
To conduct our analysis for this section, we consider \textit{fine-tuning edges}. We omit from consideration structures of model merges, adaptations and quantizations. Omitting merges allows us to work with a \textit{tree}---that is, a graph where nodes have at most one predecessor---akin to asexual reproduction.\footnote{We leave this as an open direction the genetic analysis of model merges, which can be thought of as a form of sexual reproduction with two or more parents. Graphs depicting model merges are no longer trees, and the set of local family structures is more complex. Another challenge is the frequency of model merges, which we find are considerably more rare than other forms of reproduction.} We omit adaptations and quantizations because these forms of reproduction are far less likely to branch and further support their own offspring compared to finetunes, and we intend this analysis to provide insights about the propagation of attributes over multiple generations. 

\textbf{Sampling immediate family structures.} Our aim is to measure the similarities between models residing within different `immediate family' structures in our large tree graph. One challenge with estimating quantities over these local structures is that they may appear combinatorically many times in a large graph. To illustrate what we mean by this, consider the set of all pairs of siblings in a tree graph. If one model has 500 children, the total number of pairs of siblings among them is $\binom{500}{2}$ or $124,750$. Therefore, estimating the typical similarity over all pairs of siblings quickly becomes computationally burdensome. To handle this challenge, we design an estimation procedure, where we draw a representative sample from the set of all pairs of models meeting a certain relational criterion. Our approach specifies a condition for checking whether a particular node, or a particular edge, resides within a certain structure of subgraph, and if it does, we count how many such subgraphs it belongs to. Continuing our example of counting siblings, if any node $u$ has more than one child, this condition allows us to conclude that it belongs to at least one sibling subgraph (depicted in the third row of Table \ref{tab:subgraph-sampling}). By counting $u$'s children $n_{\text{succ}}(u)$, we can infer the number of total such subgraphs $u$ contributes to, calculated as, $\binom{n_{\text{succ}}(u)}{2}$.

If we keep a lookup table of all such nodes meeting the subgraph condition, and the multiplicity of pairs contributed, then we can draw a weighted sample of the nodes in our lookup table to efficiently estimate quantities defined over sibling pairs. The full set of conditions, multiplicity relations, and subtree structures we make estimates over is depicted in Table \ref{tab:subgraph-sampling}. We sample all possible subgraphs of size 2, 3, and 4, and we estimate the similarities between all possible pairs of nodes within these subgraphs---parents, grandparents, siblings, etc.

\textbf{Family resemblance and diffusion characteristics.} 
Our main results are depicted in Figure \ref{fig:tfidf-metadata}. The results suggest that models that are close in network topology have considerably more similarity than randomly selected pairs of nodes. This offers some evidence that model family trees truly do exhibit family resemblances. However, patterns of similarity over family trees are not cleanly predicted by typical models of genetic diffusion. For example, we find that siblings are significantly more similar to one another than either is to its parent, on average (depicted in the first subfigure labeled `C'). This is counter to what an asexual model of genetic reproduction with mutation might predict. If we imagine each child model in a family inheriting the parent's genes subject to some rate of random mutation, siblings should be more related to their parent than each other, on average. We observe the opposite, suggesting that there is some directional effect of fine-tuning whereby all children tend to depart in attributes from their parents, on average, in characteristically similar ways (illustrated in Figure \ref{subfig:mutation-illustration}). 

When we look at pairs of nodes in a variety of subgraphs, we see evidence of three major heuristics that seem to dictate the level of similarity between pairs of models:
\begin{enumerate}
    \item \textbf{Same family}: If models belong to the same family tree, they appear to exhibit significantly higher levels of similarity, compared to models paired at random over our dataset.
    \item \textbf{Low generational divide}: When we compare two models that are the same \textit{generation} in their family tree (e.g., siblings or cousins), we find that this majorly increases the level of similarity between models. Models that are one generation apart (e.g., parent/child pairs or uncle/nephew pairs) tend to be significantly more similar, on average, than models that are two generations apart (e.g., grandparent/grandchild pairs). The same relationship holds when comparing grandparent/grandchild pairs to great-grandparent/great-grandchild pairs.
    \item \textbf{Network distance}: A third heuristic that seems to explain the observed similarities in model attributes is the network distance, that is, the total number of edges one would need to traverse to get from one node pair to the other. This is what a genetic model of mutation-based asexual reproduction would predict. 
    This factor is supported by the fact that uncle/nephew pairs are observed to be less similar, on average, than parent/child pairs belonging to the same subgraph structures (depicted in the second and third columns of subfigure D3 in Figure \ref{fig:tfidf-metadata}). Though most measures suggest generational divide outweighs network distance in importance, there is one exception: In the last two similarity measures in D3, we observe a parent-child pair with network distance one exhibits higher similarity than a sibling pair with network distance two.
\end{enumerate}

There are further questions that one might ask about how the diffusion of characteristics relates to the strategies and decisions ML developers make. Scientists have used models of genetic diffusion to predict cooperative, altruistic, competitive, and perhaps even spiteful behaviors in living species. For example, the theory of kin selection predicts altruistic and competitive behaviors in wolves, which are known to engage in both 1) extreme forms of self-sacrifice for their closest genetic relatives in their packs, and 2) extreme forms of competition and fighting with more distant genetic relatives in other packs \cite{cassidy2016gray}. Open ML ecosystems similarly exhibit complex dynamics with fierce forms of competition and extreme forms of cooperation. Model providers are known to compete and may undercut each other in certain ways, but still exhibit altruistic behaviors to enable third-party development, for example by releasing model weights. The analysis in this section reveals that models have more attributes in common with members of their family, and future work could explore whether these relationships predict altruistic and competitive behaviors across this ecosystem.

\section{Evolution of traits}

The previous section examined overall similarities between models across their recorded features. This section is concerned with \textit{individual traits}, focusing on their inheritance and evolution. Unless otherwise specified, we refer to categorical features as \textit{traits} and concentrate on three examples: \textit{license}, \textit{language} and \textit{task}. Through analyzing these, we demonstrate a general method for understanding trait evolution across family trees. We also report results on numerical attributes including the length of relevant documents and the number of languages.

In many cases, traits remain the same between parent and child. However, if traits were \textit{always} constant between parent and child, we'd observe far less heterogeneity in our data, and we'd find perfect similarity across all related model pairs in Figure \ref{fig:tfidf-metadata}. Because we do, in fact, observe feature diversity across models, here we focus on cases where model traits \textit{change} between a parent and a child, that is, cases where the parent has trait $i$, the child holds trait $j$, and $i\neq j$. Further information about our formal way of defining the rate of mutation is provided in Appendix \ref{app:mutation}. In observing these instances of mutation, we make a number of specific observations and findings pertaining to the individual traits in question (discussed in the proceeding subsections). More generally, we make two empirical observations that hold descriptively, but are not necessary or obvious.
\begin{enumerate}
    \item \textbf{Directedness}: We observe that mutations tend to be overwhelmingly \textit{directed}. Formally, for any two traits $(i,j)$, it is most common that $i$ overwhelmingly mutates to $j$ or that $j$ overwhelmingly mutates to $i$, rather than some balance of `traffic' of mutations in both directions. We call this phenomenon a \textit{drift}.
    \item \textbf{Orderedness}: When we consider the orientations of all directed mutations, we find that these orientations are \textit{ordered}. If we define the oriented graph of `typical' transitions between traits, we are able to find orderings over these transitions that explain virtually all these orientations.
\end{enumerate}
Notice that the first observation does not imply the second. It could be that $i$ overwhelmingly mutates to $j$, which overwhelmingly mutates to $k$, which overwhelmingly mutates back to $i$. We do not observe this for the vast majority of drifts. Second, we note that the task of finding an ordering over a directed graph is an integer programming problem, NP-hard in the worst cases. Our implementations are able to find optimal orderings, not due to luck but due to the natural orderings that emerge from our oriented graphs.

\subsection{Licenses drift from commercial to permissive and copyleft.}

How do license assignments change and mutate across model lineages?  We count 162 unique license types on the Hugging Face Hub, 98 of which are standardized license categories provided by the platform (excluding user-defined ``unknown'' and ``other'' licenses). Each model contains one license, so our reported mutation rate simply tracks the number of times a parent and child have different licenses, as a fraction of the overall number of observed inheritances.

\begin{figure}[!ht]
    \centering
    \begin{subfigure}[b]{0.58\linewidth}
        \centering
        \includegraphics[width=\linewidth]{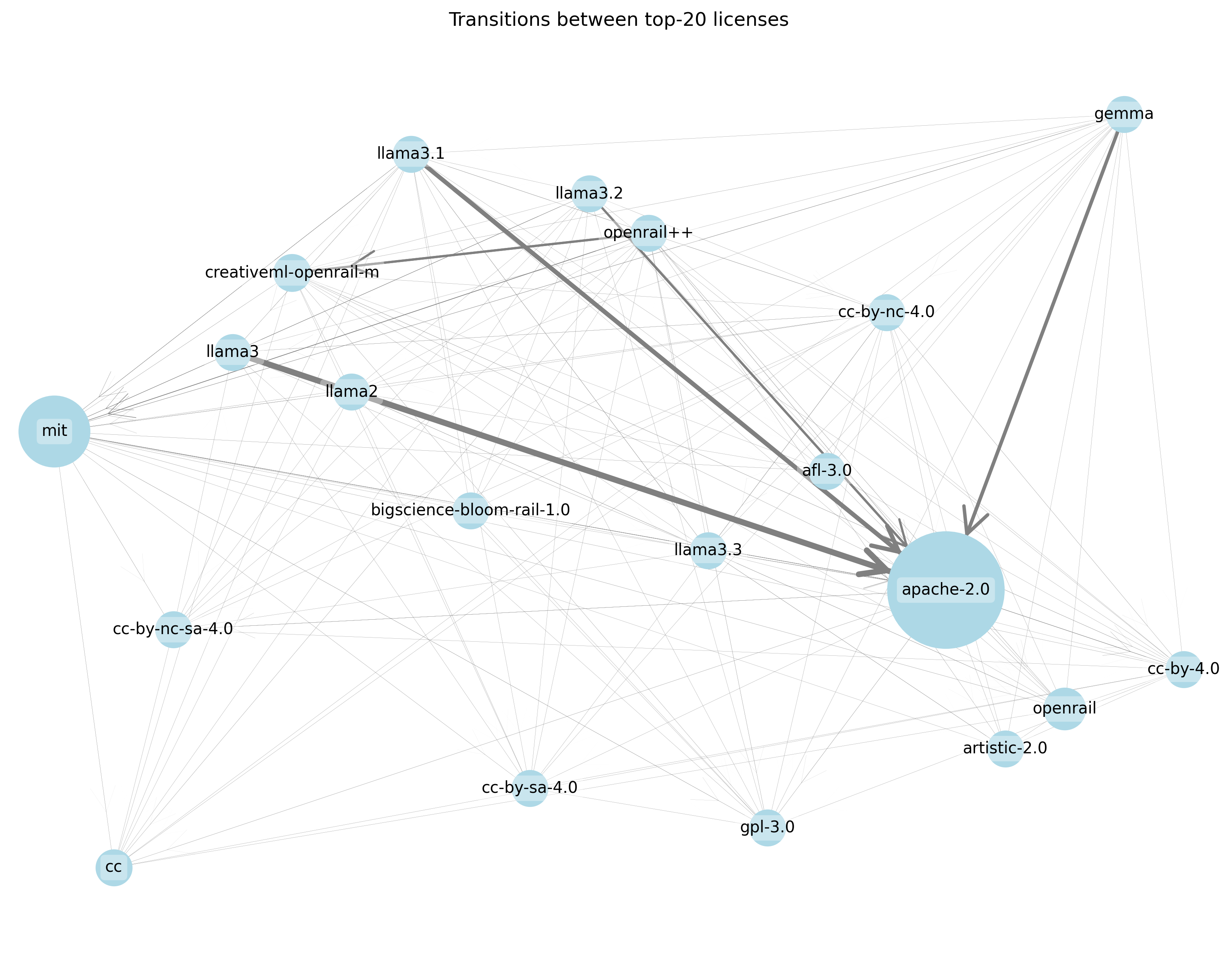}
        \label{fig:license-transitions}
    \end{subfigure}
    \hfill
    \begin{subfigure}[b]{0.38\linewidth}
        \centering
        \begin{tikzpicture}[
            mybox/.style={
                draw=black!60,
                fill=gray!8,
                rounded corners=3pt,
                line width=1.2pt,
                align=left,
                inner sep=12pt,
                text width=0.88\linewidth,
                font=\rmfamily
            },
            arrowstyle/.style={
                -{Stealth[length=3mm,width=2.5mm]},
                line width=1.5pt,
                color=black!70
            },
            thick_edge/.style={
                line width=2.5pt,
                color=black!80
            },
            medium_edge/.style={
                line width=1.8pt,
                color=black!65
            },
            thin_edge/.style={
                line width=1.0pt,
                color=black!50
            }
        ]
            \node[mybox] {
                \footnotesize\rmfamily \textbf{Optimal ordering}: \scriptsize\texttt{gemma, llama3.1, llama3.2, openrail++, creativeml-openrail-m, cc-by-nc-4.0, llama3, llama2, mit, afl-3.0, bigscience-bloom-rail-1.0, llama3.3, apache-2.0, cc-by-nc-sa-4.0, cc-by-4.0, openrail, artistic-2.0, cc-by-sa-4.0, gpl-3.0, cc} \\
                \tiny\ \\
                \footnotesize \textbf{Observed inheritances}: 320,065 \\
                \textbf{Mutation rate}: 14.98\% \\
                \textbf{Drifts following this order}: \\132/140 (94.29\%) \\
                \textbf{Mutations following this order}: 84.26\%
            };
        \end{tikzpicture}
        \label{fig:license-summary}
    \end{subfigure}
    \caption{\rmfamily An oriented directed network of license mutation drifts (left) and corresponding summary statistics (right). Each node represents a license in our dataset, with directed edges indicating mutations in which a parent and child have different licenses. Drifts seem to be directed from commercial and restrictive licenses to permissive and copyleft licenses.}
    \label{fig:transitions-graphs-license}
\end{figure}

Our analysis of the direction of evolution of licenses is summarized in Figure \ref{fig:transitions-graphs-license}. The figure depicts the most common licenses and the `drifts' between them---that is, the arrows point in the more frequent mutation direction over all observed mutations. The graph is an oriented directed graph of all 140 drifts between 20 traits, where edge weight depends on the total traffic of mutations. Using the graph, we can ask, what ordering over traits is most compatible with these drifts? If mutations were fully random, or if cycles were common, we would not be able to produce an ordering that captures more than approximately half of the observed mutation directions. However, we are able to produce an ordering accounting for 94\% of all drift directions, and 84\% of all mutations. This suggests a strong directedness in the evolution of licenses. And, equipped with this ordering, we can begin to develop hypotheses about the environmental pressures leading to the observed evolution.

The license mutations exhibit a somewhat surprising pattern. We observe many instances in which the more restrictive, commercial licenses are \textit{upstream} from the more permissive licenses.\footnote{When we refer to the categories of permissive, restrictive, commercial, and copyleft, we are using taxonomies and descriptors from existing scholarship, most notably from \citet{longpre2023data} in the context of machine learning licenses.} 
Consider, as one example, the \texttt{gemma} license, which appears first in our observed ordering. The terms of this license include the following requirement: ``You must provide all third party recipients of Gemma or Model Derivatives a copy of this Agreement.'' The license further lists use restrictions, including a restriction on uses that ``sexually explicit content, including content created for the purposes of pornography or sexual gratification (e.g. sexual chatbots).'' This license mutates most frequently to \texttt{Apache-2.0} and \texttt{MIT} licenses, each which contain no such provisions. As a second example, we observe mutation drifts from \texttt{cc-by-nc-4.0}, a ``copyleft'' license that restricts derivatives from commercial uses, to \texttt{MIT}, which grants permissions ``without limitation the rights to use, copy, modify, merge, publish, distribute, sublicense, and/or sell.'' The same non-commercial license also mutates to other licenses \textit{of the same variety} (Creative Commons) but without the non-commercial agreement, which seems to be a strict relaxation of terms.

These instances of `relaxations' appear to be the norm rather than the exception. Of the first eight licenses in our ordering, seven are commercial (\texttt{gemma} or \texttt{llama} varieties) or otherwise restrictive (\texttt{openrail} varieties). Of the last eight licenses in our ordering, none are commercial; three are permissive or public domain (\texttt{cc}, \texttt{apache-2.0}, \texttt{artistic-2.0}) and four others are copyleft varieties (\texttt{cc-by-*}, \texttt{gpl-3.0}). Looking exclusively at creative commons licenses, non-commercial restrictions lie upstream from versions without these previsions. 

Why would licenses weaken and relax even when doing so might constitute a violation of upstream agreement terms? The observed mutation drift suggests market and behavioral pressures toward openness outweigh the specter of legal enforcement as a motivator for AI developers. 

\subsection{Documentation thins.}

Armed with a method for analyzing how traits drift and evolve over generations, we now turn our attention to information about the model cards. Specifically, we are interested in the effort and resources devoted to documentation and transparency for models of different generations in the open source ecosystem. 
One significant trend that we observe is that documentation thins. Markers of bespoke effort aimed at supporting users, communicating methods, and demonstrating capabilities seem to atrophy. Markers of leaner approaches and automation develop and multiply.

\begin{figure}
    \centering
    \includegraphics[width=0.7\linewidth]{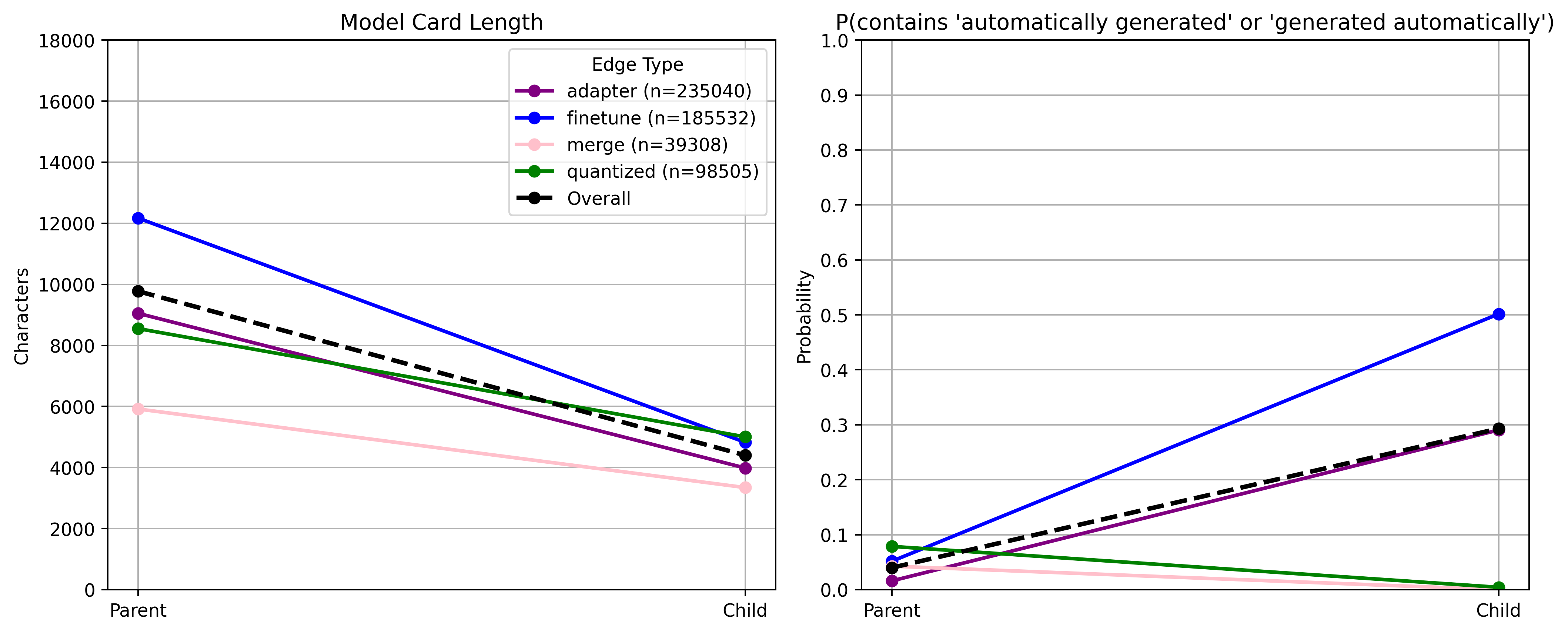}
    \caption{Even though model cards diminish in length over generations (left), the absolute frequency with which they use the terms `automatically generated' or `generated automatically' increases precipitously (right). These markers of auto-generated cards are uniquely observed in finetunes and adapters, suggesting these are byproducts of existing packages and libraries that enable developers to create finetunes and adaptations, and not those that enable merges and quantizations.}
    \label{fig:modelcardlengths}
\end{figure}

When we look at the state of model cards between parents and children in our family trees, we can make a few straight-forward observations. Model cards exist at a very high rate for models that belong to family trees. Missing model cards are far more frequent among models with no family ties. Among models with family ties, the model card is almost always available, even if it is only a few characters long. Among parent-child pairs with model cards, we observe that the length of these cards drops by $\approx 5,000$ characters. The parent's model card is roughly twice the size of the child's model card, on average. Even though the model cards get significantly shorter, we observe that they more frequently contain the terms that suggest automatic card generation. About 30\% of derivative models contain the bigrams \texttt{automatically generated} or \texttt{generated automatically}. These results, depicted in Figure \ref{fig:modelcardlengths}, suggest pressures toward lean documentation and automation technologies that remove costs to document and explain models, their capabilities, their uses, and other information typically contained in the model card.

\subsection{Languages specialize and drift toward English.}

\begin{figure} 
    \centering
    \begin{subfigure}[b]{0.45\linewidth}
        \centering
        \includegraphics[width=\linewidth]{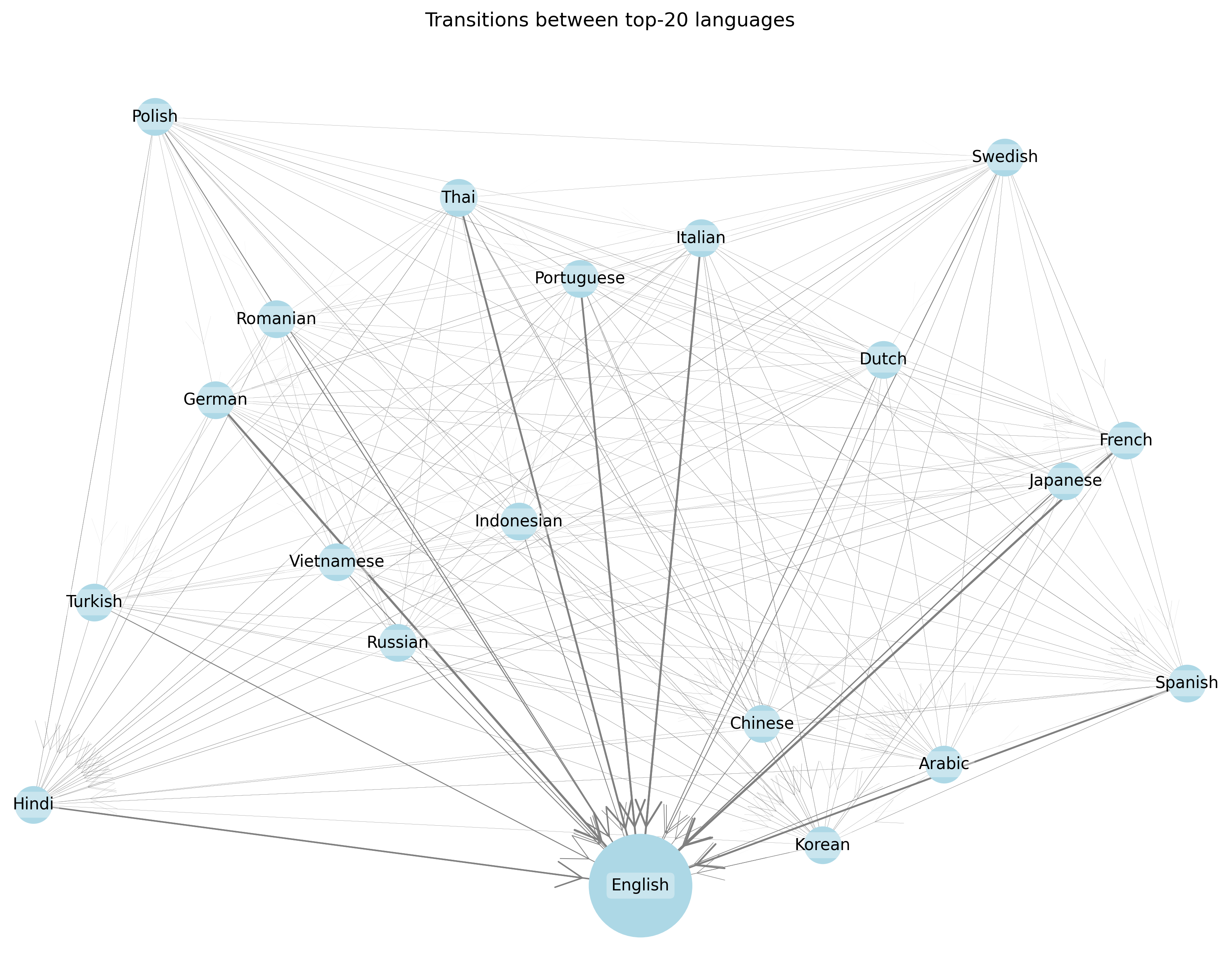}
        \label{fig:language-transitions}
    \end{subfigure}
    \begin{subfigure}[b]{0.295\linewidth}
        \centering
        \begin{tikzpicture}[
            mybox/.style={
                draw=black!60,
                fill=gray!8,
                rounded corners=3pt,
                line width=1.2pt,
                align=left,
                inner sep=12pt,
                text width=0.88\linewidth,
                font=\rmfamily
            },
            arrowstyle/.style={
                -{Stealth[length=3mm,width=2.5mm]},
                line width=1.5pt,
                color=black!70
            },
            thick_edge/.style={
                line width=2.5pt,
                color=black!80
            },
            medium_edge/.style={
                line width=1.8pt,
                color=black!65
            },
            thin_edge/.style={
                line width=1.0pt,
                color=black!50
            }
        ]
            \node[mybox] {
                \tiny\rmfamily \scriptsize\textbf{Optimal ordering}: \texttt{Polish, Swedish, Thai, Italian, Portuguese, Romanian, Dutch, German, French, Japanese, Indonesian, Vietnamese, Turkish, Russian, Spanish, Chinese, Arabic, Hindi, Korean, English} \\
                \tiny\ \\
                \scriptsize\textbf{Observed inheritances}: 115,660 \\
                \textbf{Mutation rate}: 12.80\% \\
                \textbf{Drifts following this order}:
                186/190 (97.89\%) \\
                \textbf{Mutations following this order}: 74.71\% \\
                
            };
        \end{tikzpicture}
        \label{fig:language-summary}
    \end{subfigure}
    \hfill
    \begin{subfigure}[b]{0.2125\linewidth}
        \centering
        \includegraphics[width=\linewidth]{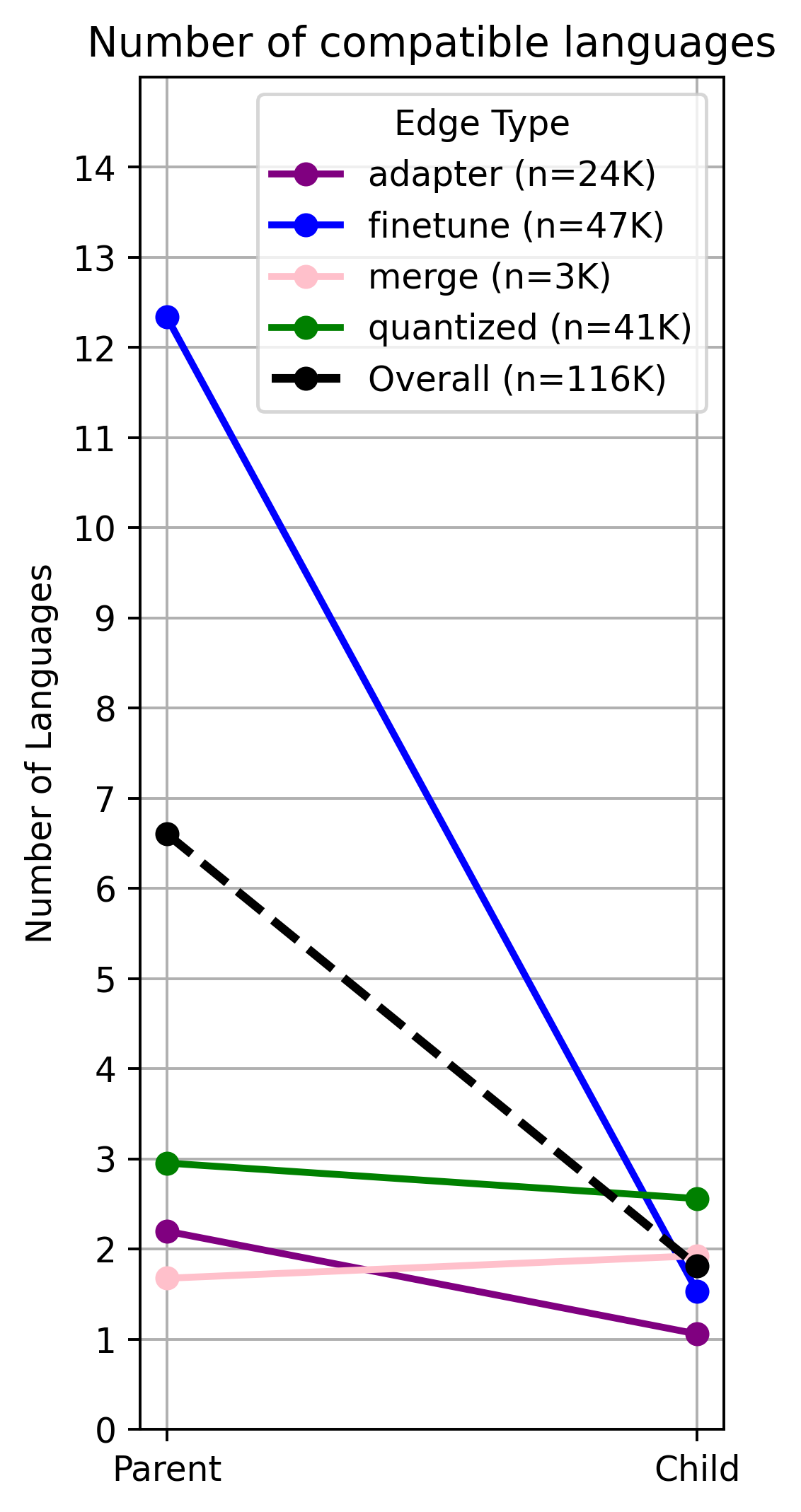}
    \end{subfigure}
    \caption{\rmfamily Network of language mutation drifts (left) and corresponding summary statistics (center). Each node represents compatibility with a language in our dataset, with directed edges indicating mutations in which a parent and child have different language compatibilities. Drifts are directed towards English. We also see considerable diminishing language support over generations (right).}
    \label{fig:transitions-graphs-language}
\end{figure}

Language traits are different in kind from licenses because an individual model can be compatible with more than one language. Whereas a mutation is a binary event in the case of license traits, for languages we allow partial mutations. Consider a case where model $i$ finetunes to model $j$. Model $i$ has language group $(A,B,C)$ and $j$ has language group $(B,C,D)$. We say that the overall mutation rate is the shared members of both groups divided by the union of both groups (i.e., in this example, the mutation rate would be $\frac{1}{2}$). Further, we log distinct directional mutations from every dropped language to every child language, and from every parent language to every added language. To continue our example, we’d log mutations from $A$ to $B$, $C$, and $D$ and from $A, B$ and $C$ to $D$. These enable us to produce similar drift diagrams and orderings to those produced for licenses. Our findings are summarized in Figure \ref{fig:transitions-graphs-language}.

The language traits show two dominant trends: 1) specialization and 2) drift towards English. The first of these trends, specialization, refers to the significant reduction in language compatibility from base models to child models. Large base models supporting significant family trees tend to support many languages, whereas derivative models tend to list compatibility with one or a handful of languages. Therefore, we see a precipitous reduction in the language support between parents and finetuned children.

The second observation we can make about language traits is that they drift overwhelmingly from broad language support to English-language support. This drift suggests a considerable market pressure towards English-speaking products and compatibilities. This drift is not entirely surprising given Hugging Face is a United States-based company. However, an increasing number of Chinese models are being developed and hosted and we do not observe a commensurate drift towards Chinese compatibility.

\subsection{Tasks appear to recapitulate the machine learning lifecycle.}

The final branch of our analysis of individual traits will focus on the \textit{tasks}. Tasks capture the model's core capability or modality: Is the model a text generation model? An image generation model? A feature extraction model? A text-to-image, image-to-text, text-to-text, or other modality-translation task? Is it a classification model? Like the model licenses, tasks typically have a one-to-one mapping to models, and so the overall rate of mutations can be interpreted simply as the fraction of edges where parent and child have different tasks. The rate of mutation is high, at 23\%, and the ordering we uncover reveals a very strong directed evolution, where 95\% of mutations are observed along the ordering.

\begin{figure*}[!ht]
    \centering
    \begin{subfigure}[b]{0.55\linewidth}
        \centering
        \includegraphics[width=\linewidth]{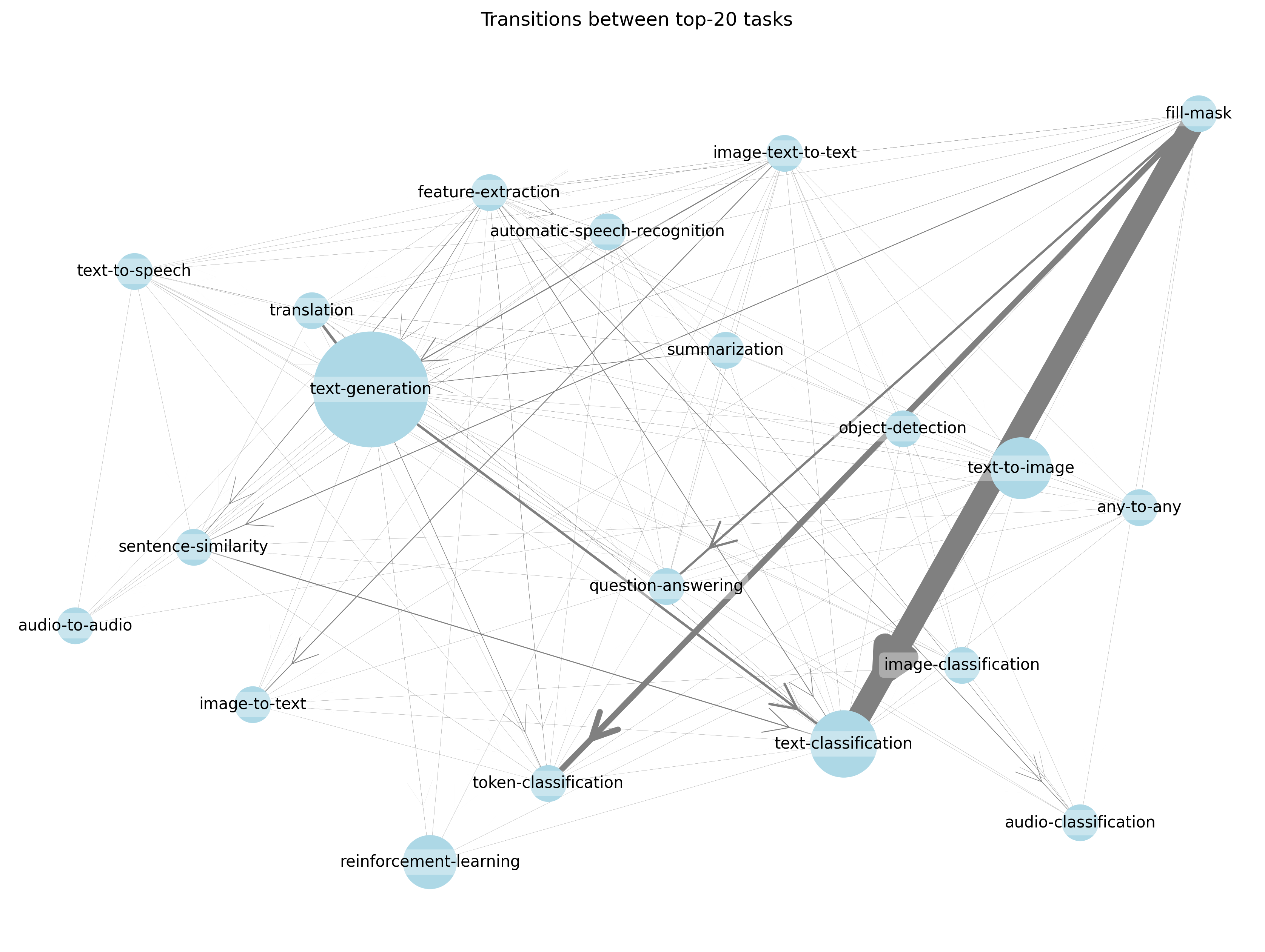}
        \label{fig:task-transitions}
    \end{subfigure}
    \hfill
    \begin{subfigure}[b]{0.44\linewidth}
        \centering
        \begin{tikzpicture}[
            mybox/.style={
                draw=black!60,
                fill=gray!8,
                rounded corners=3pt,
                line width=1.2pt,
                align=left,
                inner sep=12pt,
                text width=0.88\linewidth,
                font=\rmfamily
            },
            arrowstyle/.style={
                -{Stealth[length=3mm,width=2.5mm]},
                line width=1.5pt,
                color=black!70
            },
            thick_edge/.style={
                line width=2.5pt,
                color=black!80
            },
            medium_edge/.style={
                line width=1.8pt,
                color=black!65
            },
            thin_edge/.style={
                line width=1.0pt,
                color=black!50
            }
        ]
            \node[mybox] {
                \scriptsize\rmfamily \textbf{Optimal ordering}: \scriptsize\texttt{fill-mask},\texttt{image-text-to-text}, \texttt{feature-extraction}, \texttt{automatic-speech-recognition}, \texttt{text-to-speech}, \texttt{translation}, \texttt{summarization}, \texttt{text-generation}, \texttt{object-detection}, \texttt{text-to-image}, 
                \texttt{any-to-any}, 
                \texttt{sentence-similarity}, \texttt{question-answering}, 
                \texttt{audio-to-audio}, \texttt{image-classification}, \texttt{image-to-text},  \texttt{text-classification}, \texttt{token-classification}, 
                \texttt{audio-classification},
                \texttt{reinforcement-learning}\\
                \tiny \ \\
                \scriptsize\textbf{Observed inheritances}: 251,060 \\
                \textbf{Mutation rate}: 23.14\% \\
                \textbf{Drifts following this order}: 
                111/121 (91.74\%)\\
                \textbf{Mutations following this order}: 95.16\%
            };
        \end{tikzpicture}
        \label{fig:task-summary}
    \end{subfigure}
    \caption{\rmfamily An oriented directed network of task mutation drifts (left) and corresponding summary statistics (right). Each node represents a task in our dataset, with directed edges indicating mutations in which a parent and child have different tasks. There is a high rate of mutation, and the direction of drifts seems to recapitulate the machine learning pipeline.}
    \label{fig:modalities-graphs-task}
\end{figure*}

Our results are depicted in Figure \ref{fig:modalities-graphs-task}. The results seem to suggest that tasks progress from low-level feature extraction tasks (e.g., \texttt{fill-mask}, \texttt{feature-extraction}, \texttt{automatic\--speech\--recognition}) to modality translations (e.g., \texttt{translation, text-generation, summarization, text-to-image}), to classification and reinforcement learning tasks. 

One interpretation of this progression is that This progression seems to reflect stages in the machine learning training pipeline—beginning with general-purpose pre-training, followed by fine-tuning and alignment—mirroring how capabilities emerge and specialize over time \cite{wolf2020transformers, hub-stats}. Foundational capabilities seem to appear first, followed by modality-specific tasks, to human-aligned reasoning that produces outputs for human use.

The machine learning pipeline might be summarized as follows: First, raw inputs such as text, speech, or video are tokenized or embedded into vectors to prepare them for model processing \cite{devlin2019bertpretrainingdeepbidirectional, zhang2025qwen3embeddingadvancingtext, liu2020surveycontextualembeddings}. Second, models perform representation learning by generating contextual embeddings through masked token prediction and related techniques \cite{deepseekai2025deepseekv3technicalreport}. Third, tasks become oriented around model adaptation, such as handling cross-modal inputs or supporting low-resource languages \cite{yang-etal-2022-low}. Fourth, classification tasks assign discrete labels to how models have understood inputs \cite{kotsiantis2007supervised} (their learned representations). Fifth, generative tasks such as summarization and text-to-image synthesis produce outputs from context \cite{stiennon2022learningsummarizehumanfeedback, 10.1145/345508.345566, zhang2024texttoimagediffusionmodelsgenerative}. Lastly, techniques like instruction tuning and reinforcement learning help improve how models generate outputs, especially for tasks that require complex reasoning or alignment with human preferences \cite{longpre2023flan, dasgupta2019causal, openai2024gpt4technicalreport}.

Overall, we offer this interpretation as a possible explanatory hypothesis. We believe further research would be needed to substantiate or refute such an interpretation. Just as theories of recapitulation in human development have been contested and in many cases refuted\footnote{Recapitulation theory is largely discredited theory of development. The view is often summarized by the idea that ``ontogeny recapitulates phylogeny'' \cite{haeckel1866generelle,meckel1808beytrage,serres1859principes}.} we hope these observations are generative in motivating further inquiry. 

\section{Discussion}

In this paper we have proposed and taken a step towards studying the evolutionary biology of machine learning models. We offer a publicly-accessible dataset revealing the rich linkages between models and the genes and traits that models carry. We further propose measures to capture the \textit{genetic similarity} between models using snippets of text summarizing model attributes. Analyzing the characteristic similarities between models belonging to different local family structures, we make certain intuitive observations (e.g., that related models are more genetically related than random pairs of models) and other less-intuitive observations (e.g., that sibling pairs are considerably more genetically related than parent-child pairs). Our analysis suggests that mutation is fast and directed, and we begin to analyze the particular directional evolution of traits, focusing on licenses, languages, model card lengths and terms, and tasks. These reveal environmental pressures towards informal, lean and open attributes, especially regarding license agreements and documentations. These pressures towards openness seem to outweigh pressures to abide by upstream agreements. Further, we observe language individuation and an overwhelming drift towards english-only compatibility, though upstream models often list compatibility across many languages.

\textbf{Limitations}. Limitations to our findings include the fact that we only account for models that have logged fine-tuning relationships on Hugging Face. Many models may be related without having these relationships. For instance, models released with different numbers of paramaters are often each available as their own base model, so we do not consider \texttt{Qwen/Qwen1.5-0.5B} and \texttt{Qwen/Qwen1.5-1.8B} to be members of the same family. Though we use metadata and model card snippets as metaphors for DNA, there are other sources of semantic information we do not access. Future work may analyze model repositories' \texttt{config.JSON} files to extract architectural parameters, such as vocabulary size (inferring the training dataset size and costs), attention heads, and hidden dimensions, to reveal further attributes of models and trace how structural traits evolve across the ecosystem. Text from code repositories and even the model weights themselves could contain additional low-level semantic encodings of model properties and internals. Finally, the timescale of this analysis is limited to the lifespan of the Hugging Face platform. However, since open models predate Hugging Face, future work could extend this analysis by incorporating historical data from earlier model repositories and academic publications to capture the complete evolutionary trajectory of open source ML.

Changes to the Hub interface (e.g., available fields and auto-generated documentation) affect what developers report about their models. As Hugging Face has evolved, shifts in reporting behavior may introduce inconsistencies in the data that reflect platform changes rather than changes in the models themselves. For example, the \texttt{CreatedAt} field was introduced in March 2022 and all existent models were back-filled values equal to the date of feature launch. This could possibly inflate the genetic similarity between such models.

\textbf{Future directions}. We see the present work as a first step towards a range of studies that this dataset and perspective could support. For example, though our data represent a snapshot in which models exhibit fixed qualities, there are a variety of attributes that may be time-dependent or trends that could be uncovered with time-series data. Further, where our approach focused on open source models, there is a huge industry of closed models and these ecosystems have interesting interactions.

Structural complexity arises not only from the number of descendants but also from the introduction of merges, which combine distinct lineages---essentially ‘marrying families.’ 
Mergers between models could be viewed as a form of `sexual reproduction,' contrasting with the one-to-one parent–child mappings that this paper focuses on. As merges become more popular, the Hugging Face graph may undergo a phase transition in which nearly all nodes become connected in a single, massive connected component. Further analysis is needed to understand model merges and their effect on the ecosystem.

Lastly, future work could build on the ecological ideas in this paper, exploring concepts such as niche formation, competition, cooperation, kin selection, and succession. Work in these areas could help explain how model families grow, stabilize, or die out and move into understandings of population dynamics. 

\begin{acks}
    
\end{acks}

\bibliography{bibliography}

\section{Details on the measures of genetic similarity}
\label{app:measures}

Here we provide additional details on how we measure genetic similarity between models, and we report results across the range of measures we define.

Our Figure \ref{fig:tfidf-metadata} shows one of six ways we measure genetic similarity between models. These six methods align in the general trends and interpretations reported in the paper. Here we provide details on all six.

The measures can be divided by two \textit{targets} of similarity analysis---the metadata and model cards. On each of these pieces of text, we implement three distinct measures. One measure---Levenshtein distance---computes the total character-by-character difference. The other two---Bag-of-words (BOW) and Term Frequency-Inverse Document Frequency (TF-IDF)---measure differences using the set of n-grams in the text.

\subsection{Formal definitions}

Below we state the formal definitions of our various measures of genetic similarity. All take as input a pair of strings $s_1, s_2$ and output a measure, between 0 and 1, of similarity between them.

\begin{definition}[Cosine similarity in term frequency]
    Given two strings $(s_1, s_2)$ in a set of strings $S$, we compute the \textbf{cosine similarity in term frequency} as follows. Over all strings in $S$, produce an ordered list of the $n$ most frequently appearing terms (unigrams or bigrams). Then, for any string $s_i \in S$, define the vector $v_i \in \mathbb{R}^n$ such that every value $v_i[k]$ is the number of times the $k^{\text{th}}$ term in the list appears in $s_i$. The similarity is $\frac{v_iv_j}{||v_i||||v_j||}$.
\end{definition}

\begin{definition}[Cosine similarity in term frequency-inverse document frequency]
    Given two strings $(s_1, s_2)$ in a set of strings $S$, we compute the \textbf{cosine similarity in TF-IDF} as follows. Over all strings in $S$, produce an ordered list of the $n$ most frequently appearing terms (unigrams or bigrams). Then, for any string $s_i \in S$, define the vector $v_i \in \mathbb{R}^n$ such that every value $v_i[k]$ is the product of the number of times the $k^{\text{th}}$ term appears in $s_i$ (its term frequency) and the inverse of the fraction of documents $s \in S$ which contain the term (its inverse document frequency). The similarity is $\frac{v_i v_j}{|| v_i || || v_j ||}$.
\end{definition}

\begin{definition}[Normalized Levenshtein Similarity]
    Given two strings $(s_1, s_2)$, we define the normalized Levenshtein distance (NLD) as the minimum number of character-wise insertions, deletions, or substitutions to transform $s_1$ into $s_2$, divided by $\max(\text{length}(s_1), \text{length}(s_2))$. The \textbf{normalized Levenshtein distance} is $1-\text{NLD}$.
\end{definition}

The above definitions can be computed for a general set of strings, and we report results comparing two sets of strings specifically: The metadata, which is highly structured and recorded for every model on Hugging Face, and the model cards, which is unstructured, much more variable in length, and missing for roughly a third of all models. In the body of the text, we report results on the metadata.

\subsection{Why we prefer term frequency based similarity metrics to edit distances}

We report the TF-IDF similarities in the body of the paper, and the other similarity metrics (which match in qualitative conclusions) in the appendix. We do this for two reasons. First, we believe mutations over the metadata are more a function of differences in term-based tokens rather than character-based tokens. The difference between the snippets `\texttt{license: mit}' and `\texttt{license: gemma}' should not depend on how many letters `mit' and `gemma' share. Further, the use of traits that happen to have long names does not correspond to a further genetic distance in a meaningful. For instance, the tasks `\texttt{reinforcement-learning}' and `\texttt{fill-mask}' are not different because of the number of character deletions they require; rather they are different because they are different terms. Second, Levenshtein distance is significantly affected by the ordering of terms, such that the existance of a long tag somewhere in the middle of the string could skew the distance measure. We believe these attributes are much more a function of whether their semantic markers \textit{appear} in the metadata, and less a function of their \textit{ordering} in the metadata. This is why we prefer term frequency based measures. Finally, we choose to report the measures normalized by inverse-document frequency because it is a norm in the field, but generally we note that our qualitative insights and interpretations are consistent across the proposed measures.

\begin{figure*}[!ht]
    \centering
    \includegraphics[width=\linewidth]{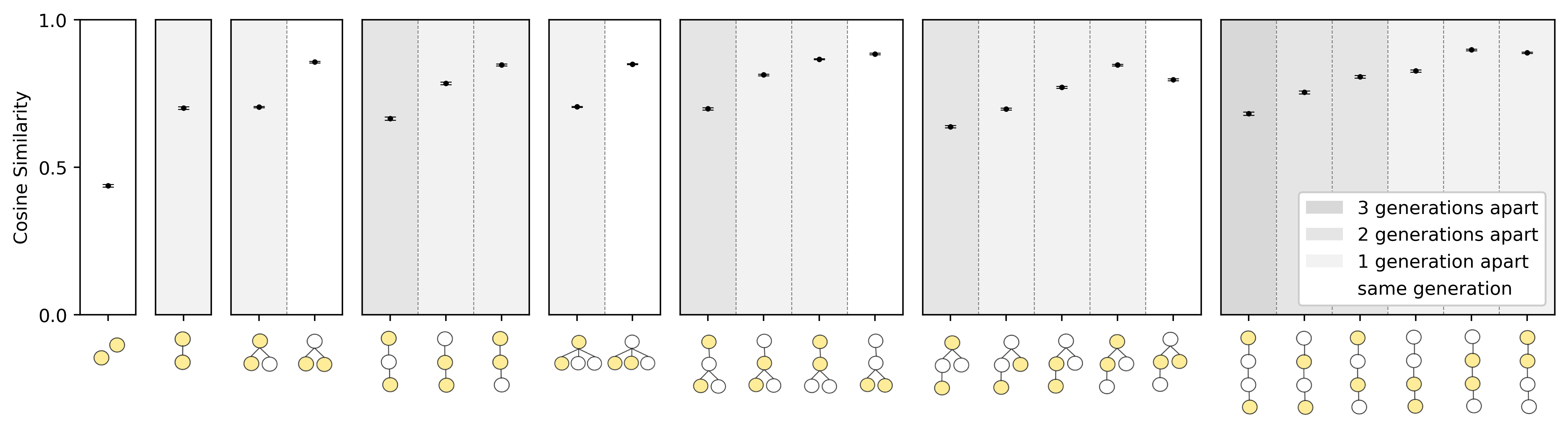}
    \caption{Bag of Words Cosine Similarity, Metadata.}
    \label{fig:bow-metadata}
\end{figure*}

\begin{figure*}[!ht]
    \centering
    \includegraphics[width=\linewidth]{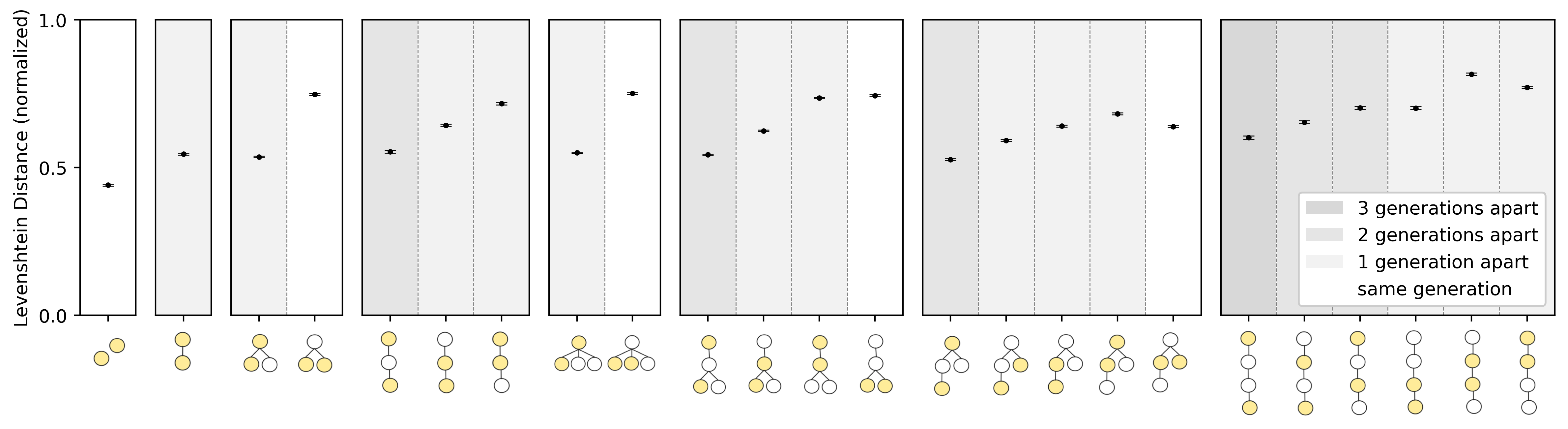}
    \caption{Levenshtein distance based similarity measure on the model metadata.}
    \label{fig:lev-metadata}
\end{figure*}

\begin{figure*}[!ht]
    \centering
    \includegraphics[width=\linewidth]{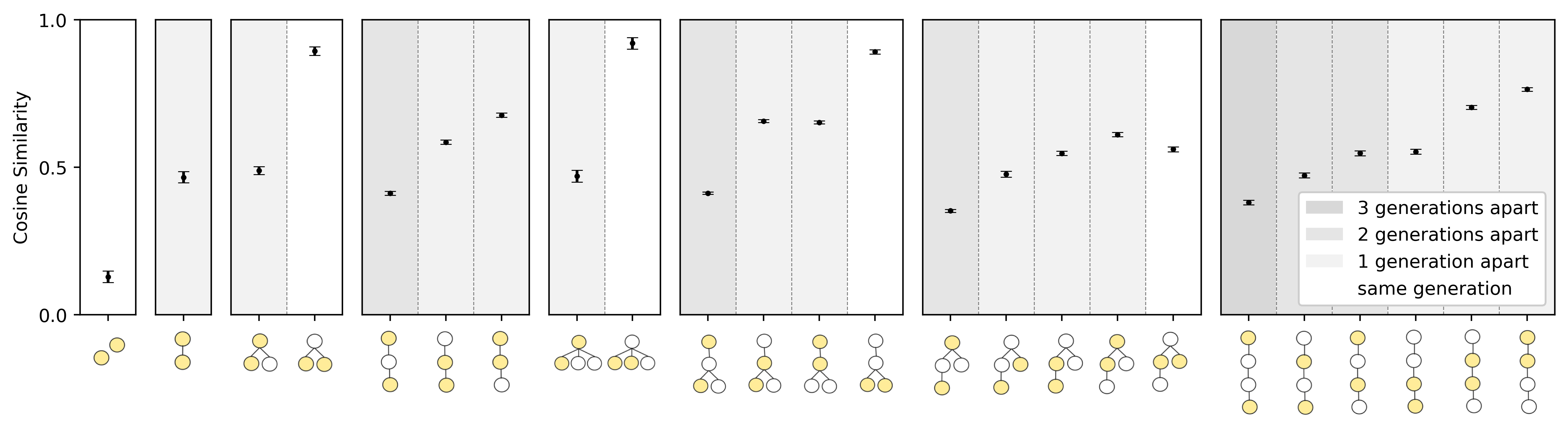}
    \caption{TF-IDF Cosine Similarity, Model Cards.}
    \label{fig:tfidf-modelcards}
\end{figure*}

\begin{figure*}[!ht]
    \centering
    \includegraphics[width=\linewidth]{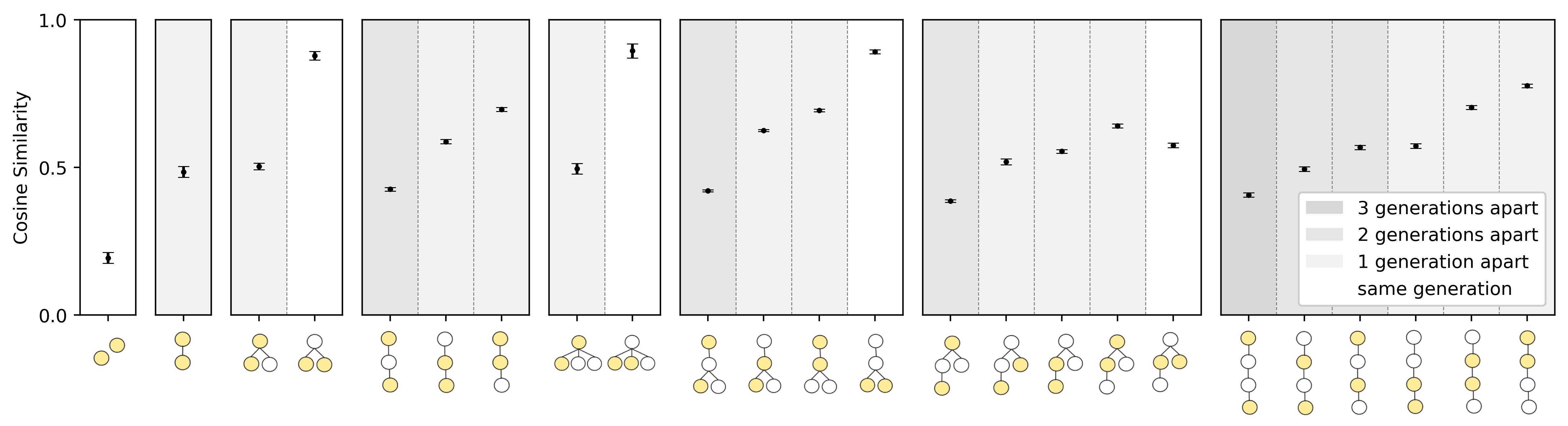}
    \caption{Bag of Words Cosine Similarity, Model Cards.}
    \label{fig:bow-modelcards}
\end{figure*}

\begin{figure*}[!ht]
    \centering
    \includegraphics[width=\linewidth]{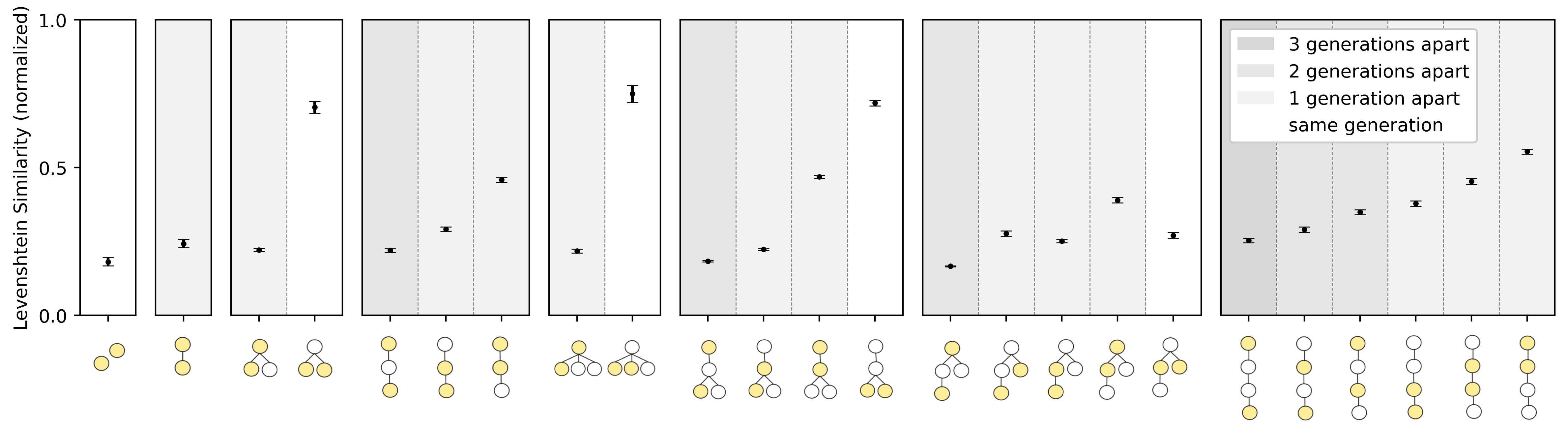}
    \caption{Levenshtein distance based similarity measure using the model cards. We have reason to believe this is the least reliable measure, as model cards are free text and Levenshtein distance relies heavily on text ordering, making it more suitable for structured strings. Even still, the directional patterns resemble the findings using other metrics.}
    \label{fig:lev-modelcards}
\end{figure*}

\section{Measuring the Mutation Rate}\label{app:mutation}

In the paper, we attempt to measure the mutation rate over model \textit{traits}. Depending on how various traits are logged in the metadata JSON, Hugging Face sometimes allows one model to list multiple traits in the same category. For other traits, however, a model can only have one categorical value. For example, models can be compatible with multiple languages, because languages are logged in the metadata as tags. Models can only have one task (or `\texttt{pipeline\_tag}'), however. Here we provide a definition for the mutation rate over a category of traits. This is the definition used in all cases where mutation rate is reported in the paper. It is compatible with both types of traits listed below (those for which models can have multiple values, and those for which models can have only one value). 

\begin{definition}[Mutation rate over traits T]
    Given a set of categorical traits $T$. Every model $i$ in our graph has a group of individual elements denoted $t_i = \{a,b, ...\} \in T$. Then the mutation rate over any directed edge $(i,j)$ is given by $m(i,j) = 1-\frac{t_i \cap t_j}{t_i \cup t_j}$. The mutation rate over the set $T$ is equal to $\frac{\sum_{\text{edges }(i,j)} m(i,j)}{N_{\text{edges}}}.$
\end{definition}
Notice that, in cases where every model must have a single categorical value in the set of traits (equivalently, $t_i$ has cardinality one $\forall i$), the mutation rate on any edge is 0 if the parent and child have the same trait, and 1 if the parent and child have different traits.

\section{Structural virality}\label{app:structural-virality}
Here, we consider the structural virality of the connected components of our dataset. Structural virality is defined as in \citet{goel2016structural}. We see evidence that structural virality is quite low in many graphs which tend to be broadcast in nature, but there nonetheless exists considerable branching, with a number of models reaching depths of nearly 40 generations.
\begin{figure*}
    \centering
    \includegraphics[width=0.9\linewidth]{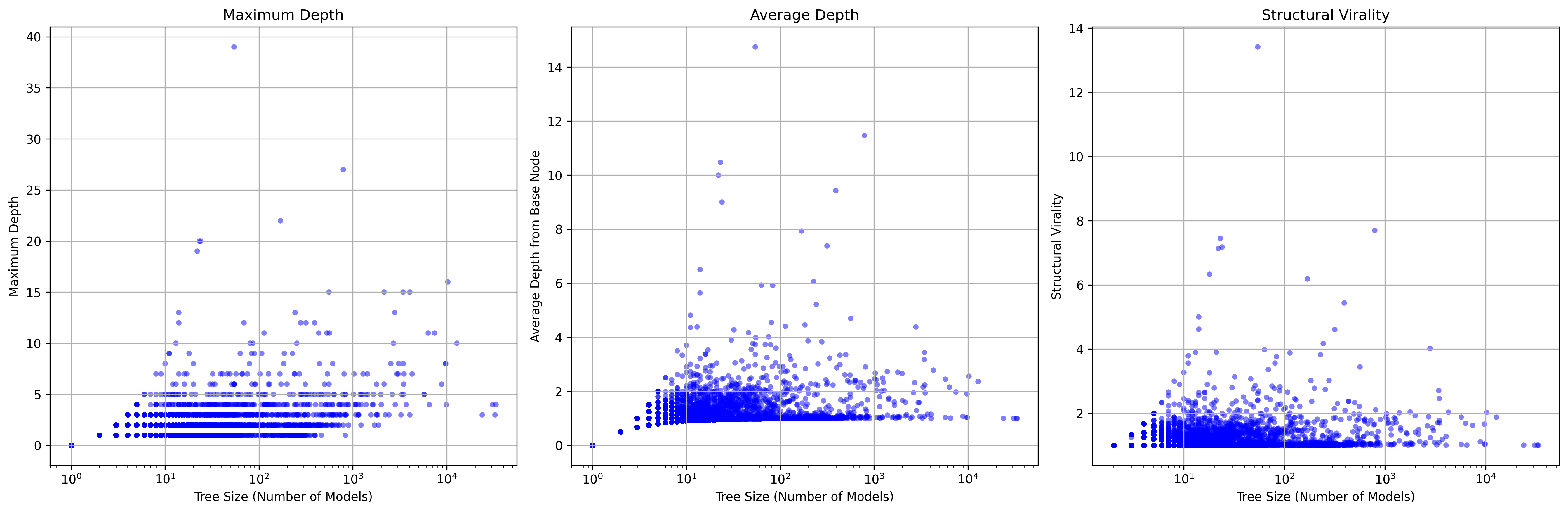}
    \caption{Scatter plots depicting the maximum depth of every node in the tree (left), the average depth of every node in the tree (center) and the structural virality (right), for trees of various sizes (measured in number of nodes). Structural virality is defined as in \cite{goel2016structural} for diffusion trees. We use only the first-listed parent for merged models to model our dataset's connected components as trees.
    }
    \label{fig:virality-analysis}
\end{figure*}

\section{Further information on the dataset}

Here we provide some additional information on the dataset and general exploratory data analysis conducted. 


\subsection{Linking papers from arXiv}  
To investigate the research inspiring models on the Hub, we extracted all linked papers from model metadata. For available arXiv IDs, we queried the \texttt{arXiv} API to retrieve the corresponding titles, abstracts, and subject classifications, allowing us to systematically categorize the papers by domain.

arXiv subject classification IDs (like cs.AI, cs.CL) are extracted from the categories column in the full JSON dataset, maps them to readable subject names using a predefined dictionary, and counts the frequency of each subject across all models. The process handles both single categories and lists of categories per model, flattening all categories into a single list before counting occurrences, where models with multiple arXiv categories contribute to the count of each individual category (e.g., a model with [`cs.AI,' `cs.CL'] adds +1 to both ``Computer Science, Artificial Intelligence'' and ``Computer Science, Computation and Language''). The top 20 most frequent research domains are then visualized in Figure \ref{fig:mega-figure}. 

\subsection{Documentation availability} We analyze model availability and observe low adoption of Hugging Face complimentary tools \cite{Goldin1996-md}.  Only 5.96\% of the models are endpoint-compatible or accessible via the Hugging Face API without local hosting. Furthermore, 6.6\% of the models released with weights use the safetensor file format---the default model weight format developed by Hugging Face in 2022 \cite{huggingfaceModelsHugging}.\footnote{Although the format was developed in 2022, it became the default (as a zero-copy alternative to \texttt{pickle}) in 2023 \cite{yoshimura2025speedingmodelloadingfastsafetensors}.} Additionally, 23.69\% of the models use automated training via Hugging Face Spaces---containerized web deployment environments. Although only a small subset of Hugging Face models have self-assigned DOIs, they are downloaded 29× more than those without. Possible explanations include DOIs make models more visible and trustworthy, and people tend to choose models that are already popular and well-documented.

\section{Further information on sampling subtree topologies}

Here we provide a more complete table as an addendum to Table \ref{tab:occurences-table}. For each shape of subgraph, we implemented a specific sampling method to get a representative sample of models. The sampling method is summarized in Table \ref{tab:subgraph-sampling}.

\begin{table*}[h!]
\centering
\begin{tabular}{>{\centering\arraybackslash}m{1.5cm} 
                >{\centering\arraybackslash}m{3.2cm} 
                >{\raggedright\arraybackslash}m{5.0cm} 
                >{\centering\arraybackslash}m{4.5cm}} 
\toprule
\textbf{Subgraph} & \textbf{
Occurrences} & \textbf{Sampling condition} & \textbf{Multiplicity$|$condition} \\
\midrule
\includegraphics[width=0.45cm]{figures/rando-icon.png} & 
3,470,193,356,870 & Two arbitrary nodes. & 1 \\
\includegraphics[width=0.25cm]{figures/duo-icon.png} & 191,072 & Single edge $(u, v)$. & 1 \\
\includegraphics[width=0.45cm]{figures/trio-1-icon.png} & 119,795,843& Node $u$ with more than one successor. & $\binom{n_{\text{succ}}(u)}{2}$ \\
\includegraphics[width=0.25cm]{figures/trio-2-icon.png} & 40,922 & Edge $(u, v)$ where $v$ has successors. & $n_{\text{succ}}(v)$ \\
\includegraphics[width=0.6cm]{figures/quad-0-icon.png} & 193,010,561,824 & Node $u$ with more than two successors. & $\binom{n_{\text{succ}}(u)}{3}$ \\
\includegraphics[width=0.45cm]{figures/quad-1-icon.png} & 11,847,103 & Edge $(u, v)$ where $v$ has more than one successor. & $\binom{n_{\text{succ}}(v)}{2}$ \\
\includegraphics[width=0.45cm]{figures/quad-2-icon.png} & 19,932,645 & Edge $(u, v)$ where $u$ has multiple successors and $v$ has successors. & $n_{\text{succ}}(v) (n_{\text{succ}}(u) - 1)$ \\
\includegraphics[width=0.25cm]{figures/quad-3-icon.png} & 10,965 & Edge $(u, v)$ where $u$ has a predecessor and $v$ has successors. & $n_{\text{succ}}(v)$ \\
\bottomrule
\end{tabular}
\caption{Subgraph patterns, their total occurrences, sampling conditions, and associated multiplicities conditioned on each pattern. $n_{\text{succ}}(u)$ refers to the number of successors (or, equivalently, the out-degree) of node $u$.}
\label{tab:subgraph-sampling}
\end{table*}

\end{document}